\documentstyle[multicol,prl,aps,amsfonts,epsfig,bm]{revtex}

\newcommand{\bra}[1]{\mbox{$\langle{#1}|$}}
\newcommand{\ket}[1]{\mbox{$|{#1}\rangle$}}

\def\beq{\begin{equation}}
\def\eeq{\end{equation}}
\def\beqa{\begin{eqnarray}}
\def\eeqa{\end{eqnarray}}
\def\I{{\rm i}}

\def\e{{\rm e}}
\def\Tr{{\rm Tr}}

\newcounter{saveeqn}
\newcommand{\alpheqn}{\setcounter{saveeqn}{\value{equation}}%
\stepcounter{saveeqn}\setcounter{equation}{0}%
\renewcommand{\theequation}{\mbox{\arabic{saveeqn}\alph{equation}}}}
\newcommand{\reseteqn}{\setcounter{equation}{\value{saveeqn}}%
\renewcommand{\theequation}{\arabic{equation}}}
\def\beql{\alpheqn \beqa}
\def\eeql{\eeqa \reseteqn}

\begin{document}

\title{
Modification of relative entropy of entanglement 
\thanks{Supported by the National Natural Science Foundation of China under Grant No. 60173047, ``973" Project of China and the National Natural Science Foundation of Anhui Province}}
\author{An Min WANG$^{1,2}$}
\address{
Laboratory of Quantum Communication and Quantum Computing
and Institute for Theoretical Physics$^1$\\  
Department of Modern Physics,
University of Science and Technology of China \\
P.O. Box 4, Hefei 230027, People's Republic of China$^2$}
\maketitle
\bigskip
\begin{abstract}
{We present the modified relative entropy of entanglement (MRE) that is proved to be a upper bound of distillable entanglement (DE), also relative entropy of entanglement (RE),  and a lower bound of entanglement of formation (EF). For a pure state, MRE is found by the requirement that MRE is equal to EF. For a mixed state, MRE is calculated by defining  a  total relative density matrix. We obtain an explicit and ``weak" closed expressions of MRE that depends on the pure state decompositions for two qubit systems and give out an algorithm to calculate MRE in principle for more qubit systems. MRE significantly improves the computability of  RE, decreases the sensitivity on the pure state decompositions in EF, reveals the particular difference of similar departure states from Bell's state  and restore the logarithmic dependence on probability of component states consistent with information theory. As examples, we calculate MRE of the mixture of Bell's states and departure states from Bell's states, and compare them with EF as well as Wootters' EF.  Moreover we study the important properties of MRE including the behavior under local general measurement (LGM) and classical communication (CC). }

\medskip
{\noindent}PACS: 03.65.Ud  03.67.-a  \vfill
\end{abstract}
\bigskip

\begin{multicols}{2}

\section{Introduction}

The entanglement is a vital feature of quantum information. It has important applications for quantum communication and quantum computation, for example, quantum teleportation \cite{Bennett1}, massive parallelism of quantum computation \cite{Ekert,QC} and quantum cryptographic schemes \cite{QCY}. Therefore, it is very essential and interesting how to measure the entanglement of quantum states. In the existing measures of entanglement, the entanglement of formation (EF) $E_{EF}$ \cite{Bennett} and the relative entropy of entanglement (RE) $E_{RE}$ \cite{Vedral1} are often used and they are respectively defined by
\vskip -0.1in
\begin{eqnarray}
E_{EF}(\rho_{AB})&=&\min_{\{p_i,\rho^i\}\in{\cal{D}}}\sum_{i} p_iS(\rho_B^i),\label{EF}\\
E_{RE}(\rho_{AB})&=&\min_{\rho^R_{AB}\in {\cal{R}}}S(\rho_{AB}\|\rho^{\rm R}_{AB}),\label{RE}
\end{eqnarray}
where ${\cal{D}}$ in eq.(\ref{EF}) is a set that includes all the possible decompositions of pure states $\rho=\sum_i p_i\rho^i$, and ${\cal{R}}$ in eq.(\ref{RE}) is a set that includes all the disentangled states. 
Note that $\rho_B^i={{\rm Tr}_A\rho^i}$ is the reduced density matrix of $\rho^i$, $S(\rho)$ is von Neumann entropy of $\rho$, $S(\rho\|\rho^{\rm R})={\rm Tr}(\rho\log\rho-\rho\log\rho^{\rm R})$ is the quantum relative entropy and $\rho^{\rm R}$ can be called the {\it relative (density) matrix}, which is used to calculate the relative entropy. 

For a pure state in a bi-party system EF is an actually standard measure of entanglement. For an arbitrary state of two qubits, EF is also widely accepted \cite{Wootters}. For bound entangled states, EF and the distillable entanglement (DE)\cite{Rains} simply quantify two different properties of the state. RE is thought of a upper bound of DE and a lower bound of EF in the case of mixed states \cite{Vedral1}. RE appears promising by a series of the interesting results \cite{Vedral2}. However, there are still several open questions not to be understood fully among them. For example, EF is heavily dependent on the pure state decompositions in the case of mixed states, RE's advantages suffers from the difficulty in computation.  Moreover, it is not very clear how to describe the entanglement of many parties in terms of both of them. At most, we can know qualitatively some useful information \cite{Vedral2}. In addition, we believe there is the particular difference between some similar departure states from Bell's state, however, it is covered up by Wootters' EF.  We do not  know why EF, in the case of mixed state, is linearly dependent on the probability of component states for the minimum pure state decomposition (MPSD). 

In this paper, we try to solve the questions stated above, at least partially. 
First, we think that in the case of pure states, EF and RE are both correct measures of quantum entanglement. Thus there must be a determined functional relation between them, but not only they are equal numerically. In other words, we should be able to find such a relative density matrix that $S(\rho\|\rho^R)=S(\rho_B)$. Although we have known that $E_{RE}(\rho)\leq E_{EF}(\rho)$ in the case of mixed states, we have no idea to find this functional relation between them. Actually, if we think that the entanglement is an inherent physical quantity of quantum state and EF and RE are both correct measures also for mixed states, then such relation definitely exists. However, EF is linearly dependent on probability of component states, but RE is logarithmically dependent on probability of component states in mathematics. It appears to  hint us that the functional relation between them might be logarithmic. Again comparing with the case of pure state, it is difficult to find a way  from a logarithmic relation to an equal relation.  This predicament is obviously an open question.  In other hand, it seems to us, EF and RE both characterize the entanglement of mixed states at a certain content. Therefore, we have to inherit their reasonable sectors and ingenious ideas. But, we also would like to improve them. 

In order to arrive at our aim, we first see what reasons lead to these difficulties. For EF, we begin with a simple example. Consider the mixed state $M$ with two kinds of pure state decompositions
\begin{eqnarray}
M&=&\frac{1}{2}\left(\ket{00}\bra{00}+\ket{11}\bra{11}\right)\\
&=&\frac{1}{4}\left(\ket{00}+\ket{11}\right)\left(\bra{00}+\bra{11}\right)\nonumber\\
& &+\frac{1}{4}\left(\ket{00}-\ket{11}\right)\left(\bra{00}-\bra{11}\right).
\end{eqnarray}
It is easy to calculate that the statistic average of EF of decomposition states are respectively  0 and 1 for two kinds of decompositions. This respectively touches at the minimum and maximum values of entanglement measure and so it is not nice enough. In order to overcome this disadvantage, one needs to find a so-called minimum pure state decomposition to define EF of a mixed state. But it appears a companying problem how to calculate the minimum pure state decomposition. At present, one seems not to know an algorithm to do this. From our view,  to calculate entanglement of the mixed states by using a minimum pure state decomposition now may be still an  indispensable trick because of the undetermined property of decomposition of density matrix. However, we can try to decrease the dependence and sensitivity with the pure decomposition so as to decrease the difficulty to find it. 

For RE, we note that  the set ${\cal{R}}$ in eq.(\ref{RE}) is so large that one can not sure when the minimization procedure is finished. In other words, although RE can measure the entanglement for bi-party systems and give out 
qualitatively description of entanglement for multi-party systems in means of the minimum distance from all of disentangled states to the concerning state, RE only pointed out that such a minimum distance exists, but does not determine what form of the disentangled state. Thus, its advantage suffers by the difficulty from computation. 

As to Wootters' EF how to cover up the difference among some departure states from the maximum entangled states can not be simply explained. We will mention it in the section four. 

Based on the definition of EF for a mixed state, we immediately see that EF is linearly dependent on probability of component states. We do not know how to  explain it from information theory. In our point of view, it seems that this dependence should be logarithmic. In fact, this is one of main reasons why we take the relative entropy to describe the measure of entanglement. However, we have to face to a new difficulty how to calculate it. 

After these analyses stated above, we realize that it is necessary and important to further  research measures of quantum entanglement. In order to restore the logarithmically dependence on probability of component states, we prefer to chose the relative entropy, as a function of mixed state, to describe the entanglement of mixed state. However, since the facts that the pure state decomposition of a mixed state is not unique in general and any decomposition is not always corresponding to the really physical entanglement, we have to  determine a pure decomposition so as to the relative entropy calculated by it can correctly measure entanglement.  In spite of the puzzle of the linearly dependence on probability of component states from EF in a mixed state,  it seems to us, the kernel of Bennett {\it et.\ al}'s idea is to point out the minimum pure state decomposition of a mixed state corresponds to the entanglement of this mixed state.  Thus,  we define MRE just according with this kernel of their idea. Moreover, in order to overcome RE's difficulty in seeking a suitable relative density matrix among an infinite set of disentangled states, we derive out an explicit  construction of relative density matrix in MRE.  In summary, the main ideas to propose MRE are original from organically combining the advantages of EF for the pure states and strongpoint of RE for the mixed states and avoiding their individual shortcomings as possibly. Of course, we have used some our points of view and judgements.  

It must be emphasized that in the case of pure states, MRE gives the same value of entanglement as EF. Therefore, MRE can be thought of an acceptable measure of entanglement in the case of pure state. In other words,  MRE's and EF's  positions are equal, that is, there is no any difference whether by means of MRE or EF  to measure entanglement of a pure state. As is well known, in the case of mixed states, EF has be extensively researched. In this paper, we would like to study how MRE to be extended.  Our results implies that MRE is indeed a hopeful candidate to measure the entanglement for the mixed states.   

Obviously, the most important key is how to construct a correct relative density matrix in MRE. Our method can be simply described as following. First,  starting with a pure state $\rho_{A\!B}^{\rm P}$,   we think the measure of entanglement is proportional to such a relative entropy $S(\rho_{A\!B}^{\rm P}||R)$,  in which the relative density matrix is defined by equation $S(\rho_{A\!B}||R)=E_{EF}(\rho_{A\!B})$ based on the fact that EF is a good enough measure of entanglement for the pure states, that is, $R$ is a solution of this equation. Then, we define the relative density matrix in means of introducing the bases of relative density matrix. In the case of mixed states, for each pure state decomposition,  we can construct an individual relative density matrix in terms of a mixture of relative density matrices of all component states with same distribution. In general,  for all of possible pure state decompositions, their corresponding relative density matrices are not the same and forms a set. Thus, among this set we, according to Bennett {\it et. \ al}'s idea, chose such a relative density matrix that  the relative entropy of mixed state evaluated by it is the minimum as a correct total relative density matrix in MRE.  Just because the relative density matrix can be constructed obviously in MRE, one can easily calculate the minimum distance and clearly understand its physical meaning.  

Of course, the simplest case is two qubits as bi-party systems. It is a footstone to understand and calculate MRE in the cases of many qubits and multi-party systems. In this paper, at least for bi-party systems made up of two qubits, we clearly derive out the forms of relative density matrices, explicitly obtain their closed expressions. All of this greatly improves the computability of relative entropy as a measure of entanglement, decreases at some content  the undetermined property of measure of entanglement of mixed states and overcomes above difficulties that we have realized.  Moreover, it is proved to be a possible upper bound of RE, also DE, and a lower bound of EF. In particular, MRE has some expected behaviors under local general measurement (LGM) and classical communication (CC). It seems to us, the advantages of  MRE might be more important for multi-party systems, and we have further developed our study to the relevant problems \cite{My1}.  More details will be presented soon. 

This paper is organized as following. Section one, as  introduction, mainly analyses the actuality and problems at front of us in the study of quantum entanglement and explains why and how to propose MRE. Section two, as preliminaries,  contains  several lemmas which are the computing method of relative entropy, physical significance and expression of polarized vectors related with entanglement, the behavior and properties of polarized vectors and disentangled states under local general measurement (LGM) and classical communication (CC). Section three proposes the full definition of MRE,  obtains a ``weak" closed expression of MRE that depends on the pure state decompositions for two qubit systems,  gives out an algorithm to calculate MRE in principle for more qubit systems. Section four exhibits some useful examples to account for the advantages of MRE including significant improvement of the computability of  relative entropy of entanglement (RE), decreasing dependence and sensitivity on the pure state decompositions  and correct  logarithmic dependence, in the sense of information theory, on probability of component states, as well as the particular difference among the departure states from Bell's states.  MRE of the mixture of Bell's states and the departure states from Bell's states are calculated and is compared with their EF as well as Wootters' EF.  Section five proves important properties of MRE such as that MRE is a possible upper bound of RE, also DE, and a lower bound of EF,  MRE has some expected behaviors under local general measurement (LGM) and classical communication (CC),  MRE varies from 0 to 1 as well as its maximum value corresponds to maximally entangled states and its minimum value corresponds to separable states.  

\section{Several Lemmas} 

As preliminaries, let's first give out the following several lemmas. 
In order to calculate relative entropy, we need

{\bf Lemma One}.\ If the relative density matrix in its eigenvector decomposition is:
\begin{equation}
\rho^{\rm R}=\sum_{\alpha}\lambda_\alpha\rho_\alpha^{\rm R}=\sum_{\alpha}\lambda_\alpha\ket{v_\alpha^{\rm R}}\bra{v_\alpha^{\rm R}},\label{ED}
\end{equation}
where $\lambda_\alpha$ is taken over all the eigenvalues and the eigen density matrices are assumed to be orthogonal and idempotent without loss of generality,  
Thus, the relative entropy can be written as
\begin{eqnarray}
S(\rho\|\rho^{\rm R})&=&-S(\rho)-\sum_{\alpha}\log \lambda_\alpha {\rm Tr}(\rho\rho_\alpha^{\rm R})\\
&=&-S(\rho)-\sum_{\alpha}\log \lambda_\alpha \bra{v_\alpha^{\rm R}}\rho\ket{v_\alpha^{\rm R}}.\label{CRE}
\end{eqnarray}

It is easy to prove lemma one by the simple and standard computation in quantum mechanics. So, we omit it. This lemma implies that the key to calculate RE is to seek an appropriate relative density matrix $\rho^{\rm R}$ and to find out all of its eigenvalues and eigenvectors.  In the construction of relative density matrix $\rho^{\rm R}$ for pure states, we will find that it  is directly related with the polarized vectors of reduced density matrices. 
For simplicity, consider the case for two qubits and denote the reduced density matrices for a quantum state $\rho$ are
\beq
\rho_A=\Tr_B\rho;\quad \rho_B=\Tr_A\rho.
\eeq
They can be rewritten as
\beq
\rho_A=\frac{1}{2}(\sigma_0+\bm{\xi}_A\cdot\bm{\sigma}), \quad 
\rho_B=\frac{1}{2}(\sigma_0+\bm{\xi}_B\cdot\bm{\sigma}), 
\eeq
where $\sigma_0$ is the identity matrix and $\bm{\sigma}$ is usual Pauli spin matrix. $\bm{\xi}_A$ and $\bm{\xi}_B$ are just polarized vectors respectively corresponding to $\rho_A$ and $\rho_B$. We always can expand the density matrices as
\begin{equation}\label{2qrho}
\rho=\frac{1}{4}\sum_{\mu,\nu=0}^{3} a_{\mu\nu}\sigma_\mu\otimes\sigma_\nu.
\end{equation}
Obviously, we have 
\beqa
\label{xiA}
\xi_A^i=\Tr(\rho_A\sigma_i)=\frac{1}{2}\sum_{\mu=0}^3\Tr(a_{\mu 0}\sigma_\mu\sigma_i)=a_{i0},\\
\label{xiB}
\xi_B^j=\Tr(\rho_B\sigma_j)=\frac{1}{2}\sum_{\mu=0}^3 \Tr(a_{0\nu}\sigma_\nu\sigma_j)=a_{0j}.
\eeqa
In general, they are not equal.  But in the case of a pure state 
\beq
\label{2qpureS}
\ket{\psi}=a\ket{00}+b\ket{01}+c\ket{10}+d\ket{11},
\eeq
it follows that
\begin{equation}
\bm{\xi}^2=\bm{\xi}^2_A=\bm{\xi}^2_B=1-4|ad-bc|^2.
\end{equation}
that is that the norms of $\bm{\xi}_A$ and $\bm{\xi}_B$ are equal.
For arbitrary quantum states, it is easy to prove that
\begin{equation}
\bm{\xi}_A={\rm Tr}(\rho \bm{\sigma}\otimes I),\quad\bm{\xi}_B={\rm Tr}(\rho I\otimes\bm{\sigma}).
\end{equation}
The relations between their components are given out in lemma two.

{\bf Lemma Two}.\  For the pure state of two qubits, there are the relations between  the polarized vectors $\bm{\xi}_{A}$ and $\bm{\xi}_B$: 
\begin{equation}
\xi^i_A=\sum_{j=1}^3a_{ij}\xi^j_B,\quad \sum_{i=1}^3\xi^i_Aa_{ij}=\xi^j_B.
\end{equation}

\noindent{\bf Proof}\ Obviously, for a pure state
\beq
\rho^2=\rho. \label{e3}
\eeq
Thus, 
\beq
\Tr\rho^2=\Tr\rho=1\label{eq18},\quad \Tr_B\rho^2=\Tr_B\rho.\label{eq19}
\eeq
Substituting eq.(\ref{2qrho}) to eq. (\ref{eq18}) and using the relations
\beql
(A\otimes B) (C\otimes D)&=&(AC)\otimes(BD),\\
\Tr(A\otimes B)&=&\Tr A\Tr B,
\eeql
we have 
\beqa
\Tr\rho^2&=&\frac{1}{16}\sum_{\mu,\nu=0}^3\sum_{\mu^\prime,\nu^\prime=0}^3 \Tr(\sigma_\mu\sigma_{\mu^\prime})\Tr(\sigma_\nu\sigma_{\nu^\prime})a_{\mu\nu}a_{\mu^\prime\nu^\prime}\nonumber\\
&=&1.
\eeqa
Because
\beqa
\Tr\sigma_\mu\sigma_\nu &=&2\delta_{\mu\nu}\quad (\mu,\nu=0,1,2,3),\\ \Tr\sigma_i&=&0\quad (i=1,2,3),\label{e6}
\eeqa
then
\beq
\frac{1}{4}\sum_{\mu,\nu=0}^3a_{\mu\nu}a_{\mu\nu}=1.\label{eq22}
\eeq
Again substitute eq.(\ref{2qrho}) to eq.(\ref{eq19}), we have
\beqa
\label{eq23}
\Tr_B\rho&=&\frac{1}{2}\sum_{\mu=0}^3 a_{\mu0}\sigma_\mu
=\Tr_B\rho^2\nonumber\\
&=&\frac{1}{16}\sum_{\mu,\nu=0}^3\sum_{\mu^\prime,\nu^\prime=0}^3 (\sigma_\mu\sigma_{\mu^\prime})\Tr(\sigma_\nu\sigma_{\nu^\prime})a_{\mu\nu}a_{\mu^\prime\nu^\prime}\nonumber\\
&=&\frac{1}{8}\sum_{\mu,\nu,\mu^\prime=0}^3 a_{\mu\nu}a_{\mu^\prime\nu}(\sigma_\mu\sigma_{\mu^\prime}).
\eeqa
Further, let's rewrite the right side in the above equation
\beqa
&&\frac{1}{8}\sum_{\mu,\nu,\mu^\prime=0}^3a_{\mu\nu}a_{\mu^\prime\nu}(\sigma_\mu\sigma_{\mu^\prime})\nonumber\\ &=&\frac{1}{8}\sum_{\nu=0}^3\left[\sum_{\mu^\prime=0}^3a_{0\nu}a_{\mu^\prime\nu} \sigma_{\mu^\prime}+\sum_{i=1}^3\sum_{\mu^\prime=0}^3a_{i\nu}a_{\mu^\prime\nu} \sigma_i\sigma_{\mu^\prime}\right]\nonumber\\
&=& \frac{1}{8}\sum_{\nu=0}^3\left[ a_{0\nu}a_{0\nu}\sigma_0+\sum_{i=1}^3a_{0\nu}a_{i\nu}\sigma_i\right.\nonumber\\
& &\left.+\sum_{i=1}^3a_{i\nu}a_{0\nu}\sigma_i+\sum_{i,j=1}^3a_{i\nu}a_{j\nu}\sigma_i\sigma_j\right] \nonumber\\
&=&\frac{1}{8}\sum_{\nu=0}^3\left[ (a_{0\nu}a_{0\nu}+\sum_{i=1}^3a_{i\nu}a_{i\nu})\sigma_0+2\sum_{i=1}^3a_{0\nu}a_{i\nu}\sigma_i\right]\nonumber\\
&=&\frac{1}{8}\left[\sum_{\mu,\nu=1}^3 a_{\mu\nu}a_{\mu\nu}\sigma_0+2\sum_{i=1}^3\left(a_{i0}+\sum_{j=1}^3a_{0j}a_{ij}\right)\sigma_i\right],
\eeqa
where we have used
\beq
\sigma_i\sigma_j+\sigma_j\sigma_i=2\delta_{ij}\sigma_0,\quad \Tr\rho=a_{00}=1.
\eeq
Then,  from eq.(\ref{eq22}) and eq.(\ref{eq23})  it follows that
\beqa
&&\frac{1}{2}\left[\sigma_0+\frac{1}{2}\sum_{i=1}^3\left(a_{i0}+\sum_{j=1}^3a_{0j}a_{ij}\right)\sigma_i\right]\nonumber\\
&=&\frac{1}{2}\left(\sigma_0+\sum_{i=1}^3a_{i0}\sigma_i\right).
\eeqa
Multiplying $\sigma_k$ to two sides and tracing it, we obtain
\beq
a_{i0}=\sum_{j=1}^3 a_{ij}a_{0j}.\label{e28}
\eeq
Likewise, in terms of $\Tr_A\rho^2=\Tr_A\rho$ we also can prove 
\beq
a_{0j}=\sum_{i=1}^3a_{i0}a_{ij}.\label{e29}
\eeq
Eqs. (\ref{e28}) and (\ref{e29}) are the relations between $a_{i0}$ and $a_{0j}$. They are not equal in general. 
Again substitute eqs.(\ref{xiA}) and (\ref{xiB}) to eqs.(\ref{e28}) and  (\ref{e29}), our lemma two is proved. 

It is useful to research the relation between entanglement and the polarized vectors.  This is lemma three.

{\bf Lemma Three}\ For a pure state of two qubits,  the entanglement is a  monotone decreasing function of  $\bm{\xi}^2$ which is the norm of the polarized vector of reduced density matrix. If  $\bm{\xi}^2=1$, it is a separable state. If  $\bm{\xi}^2=0$, it is a maximally entangled state.

Its proof is also easy. In fact,  we can calculate out that Wootter's concurrence is equal to $C=2|ad-bc|$ for a pure state (\ref{2qpureS}).  It is well known that the entanglement of a pure state for tow qubit system is monotonically increases with $C$ \cite{Wootters}.  Note that there is a relation $C^2=1-\bm{\xi}^2$, we obtain the conclusion that the entanglement is a  monotone decreasing function of the norm of the polarized vector. In special,  if $|\bm{\xi}|=1$, the reduced density matrix only has a non zero eigenvalue.  In other words, it is a pure state. Then, von Neumann entropy of reduced density matrix is zero. It implies that the corresponding pure state is  separable. While $|\bm{\xi}|=0$,  two eigenvalues of reduced density matrix are both $1/2$, Then, von Neumann entropy of reduced density matrix is 1. This corresponds to the maximally entangled states. Furthermore, we can prove immediately that the necessary and sufficient condition of a separable state is $|ad-bc|=0$, and the necessary and sufficient condition of Bell states  is $|ad-bc|=1/2$ \cite{My2}. In our point of view,  the norm of polarized vector is a simple and useful measure of entanglement in the case of pure states.  

To research the entanglement purification and distilling, we need to know behavior of entanglement under $LGM$ and $CC$. Here, $LGM+CC$ means that two parties $A$ and $B$ perform separately two sets of operations which are described by 
\begin{equation}
\rho_{AB}^{\prime\prime}=\sum_\lambda A_\lambda\otimes B_\lambda\rho_{AB}A_\lambda^\dagger\otimes B_\lambda^\dagger,
\end{equation}
where these two sets of operators satisfy the completeness relations
\begin{equation}
\sum_\lambda A_\lambda^\dagger A_\lambda\otimes B_\lambda^\dagger B_\lambda=1.
\label{TPC}
\end{equation}
While only there is one member in the above sets, it is called pure $LGM+CC$. If $A_\lambda$ or $B_\lambda$ is a unit matrix, it will belong to $LGM$. In discussion on the properties of MRE, the following lemmas are useful. 

{\bf Lemma Four}\ Under $LGM+CC$, that is,  for a pure state under the following transformation
\beq
\rho_{A\!B\lambda}^{\prime\prime}=(A_\lambda\otimes B_\lambda)\rho_{A\!B}(A_\lambda^\dagger\otimes B_\lambda^\dagger)/q_\lambda,
\eeq
the norm of transformed polarized vector $\bm{\xi}_{\lambda}^{\prime\prime\;2}$ becomes
\begin{equation}
\bm{\xi}_{\lambda}^{\prime\prime\;2}=1-\frac{4|ad-bc|^2\det(A_\lambda^\dagger A_\lambda B_\lambda^\dagger B_\lambda)}{q^2_\lambda},\label{xipp}\label{pspvpp}
\end{equation}
where $q_\lambda$ reads 
\beqa
\label{Trrholambdapp}
q_\lambda&=&\Tr[(A_\lambda\otimes B_\lambda)\rho_{A\!B}(A_\lambda^\dagger\otimes B_\lambda^\dagger)]\nonumber\\
&=&
|a^{\lambda\;\prime\prime}|^2+|b^{\lambda\;\prime\prime}|^2+|c^{\lambda\;\prime\prime}|^2+|d^{\lambda\;\prime\prime}|^2, \label{P}
\eeqa
while $a^{\prime\prime},b^{\prime\prime},c^{\prime\prime},d^{\prime\prime}$ are coefficients in the transformed state vector
\beqa
\ket{\psi^{\prime\prime}_\lambda}&=&A_\lambda\otimes B_\lambda\ket{\psi}\nonumber\\
&=&a^{\lambda\;\prime\prime}\ket{00}+b^{\lambda\;\prime\prime}\ket{01}+c^{\lambda\;\prime\prime}\ket{10}+d^{\lambda\;\prime\prime}\ket{11}.
\eeqa
which has not been normalized. 

\noindent{\bf Proof}:\  In order to prove this lemma, let's first consider the pure $LGM$ quantum operation $I\otimes B$ and denote
\begin{equation}
\ket{\psi^\prime}=I\otimes B\ket{\psi}= a^\prime\ket{00}+b^\prime\ket{01}+c^\prime\ket{10}+d^\prime\ket{11},
\end{equation}
we have then
\beql
a^\prime=aB_{11}+bB_{12},\\
b^\prime=aB_{21}+bB_{22},\\
c^\prime=cB_{11}+dB_{12},\\
d^\prime=cB_{21}+dB_{22},
\eeql
where $B_{ij}  (i,j=1,2)$ are matrix elements of $B$, so that 
\begin{equation}
a^\prime d^\prime-b^\prime c^\prime=(ad-bc)\det B.
\end{equation}
Similarly we can treat with the pure $LGM$ quantum operation $A\otimes I$. In terms of $A\otimes B=(A\otimes I)(I\otimes B)$, we arrive at
\begin{equation}
a^{\prime\prime} d^{\prime\prime}-b^{\prime\prime} c^{\prime\prime}=(ad-bc)\det A\det B.
\end{equation}
Finally, since $\ket{\psi^{\prime\prime}}$ is also a pure state, normalizing  $\ket{\psi^{\prime\prime}}$ and then using the expression of norm of polarized  vector, we immediately can obtain eq.(\ref{pspvpp}). 

{\bf Lemma Five}\  Under $LGM+CC$ quantum operation, if  $A^\dagger_\lambda A_\lambda\otimes B_\lambda^\dagger B_\lambda$ is proportional to an identity matrix, it does not change the norms of polarized vectors of reduced density matrix of a pure state and does not change the general entanglement of formation either.    

\noindent{\bf Proof} Actually,  based on the property of matrix direct product , it follows that
\beq
[\det(A^\dagger_\lambda A_\lambda B^\dagger_\lambda B_\lambda)]^2=\det(A^\dagger_\lambda A_\lambda\otimes B^\dagger_\lambda B_\lambda).
\eeq
Since eq.(\ref{Trrholambdapp}) and noting that $A^\dagger_\lambda A_\lambda\otimes B^\dagger_\lambda B_\lambda$ is proportional to an identity matrix, we have  then
\beq
\det(A^\dagger_\lambda A_\lambda B^\dagger_\lambda B_\lambda)=q_\lambda^2,
\eeq
so that
\begin{equation}
\bm{\xi}^{\prime\prime\;2}=1-4|ad-bc|^2\frac{\det(A_\lambda^\dagger A_\lambda B_\lambda^\dagger B_\lambda)}{q^2_\lambda}=\bm{\xi}^2.
\end{equation}
It indicates that the norms of polarized vectors are invariant under this transformation. In special, for a pure $LGM+CC$, since  $A^\dagger A\otimes B^\dagger B=1$ ( This is a trace preserving condition), we have the same result. Because that $|\bm{\xi}|$ can be thought of as a concurrence of EF for a pure state,  this result implies that EF is unchanged. In the case of mixed state, for each $\bm{\xi}_i$ from the component state $\rho_i$, we have the similar proof and then the same conclusions. However, for a transformation without the condition that $A^\dagger_\lambda A_\lambda\otimes B_\lambda^\dagger B_\lambda$ is proportional to an identity matrix, the norm of polarized vector changes according to eq.(\ref{pspvpp}) in general. 

{\bf Lemma Six}\ Any $LGM+CC$ can not change a unentangled state to an entangled state for the system of two qubits (Note that the measures of entanglement are always larger than or equal to 0).

\noindent{\bf Proof.}\ In general,  a pure state will transform to a mixed state under $LGM+CC$:
\beq
\rho_{A\!B}^{\prime\prime}=\sum_{\lambda}(A_\lambda\otimes B_\lambda)\rho_{A\!B}(A_\lambda^\dagger\otimes B_\lambda^\dagger)
=\sum_{\lambda}q_\lambda\rho_{A\!B\lambda}^{\prime\prime},
\eeq
where
\beqa
q_\lambda&=&\Tr[(A_\lambda\otimes B_\lambda)\rho_{A\!B}(A_\lambda^\dagger\otimes B_\lambda^\dagger)],\\
\rho_{A\!B\lambda}^{\prime\prime}&=&(A_\lambda\otimes B_\lambda)\rho_{A\!B}(A_\lambda^\dagger\otimes B_\lambda^\dagger)/q^2_\lambda.
\eeqa
Because for a unentangled state, $|ad-bc|=0$. Again from eq.(\ref{pspvpp}), it follows that $|\bm{\xi}_\lambda^{\prime\prime}|=1$. This implies that every component state $\rho_{AB\lambda0}^{\prime\prime}$ is separable. Of course,  the entanglement of transformed states is then equal to zero. For the mixed state of various unentangled states, the proof is similar. 
For example, for a separable state
\beq
\rho_{\rm S}=\sum_ip_i \rho_A^i\otimes\rho_B^i=\sum_ip_i \ket{\psi^i_{\rm S}}\bra{\psi^i_{\rm S}}.
\eeq
Obviously, because $ \ket{\psi^i_{\rm S}}\bra{\psi^i_{\rm S}}=\rho_A^i\otimes\rho_B^i=\rho^i_{\rm S}$, we can write 
\begin{equation}
\ket{\psi^i_{\rm S}}=(a_1^i\ket{0}+b_1^i\ket{1})\otimes(a_2^i\ket{0}+b_2^i\ket{1}).
\end{equation}
Comparison it with the pure state $\ket{\psi^i}=a^i\ket{00}+b^i\ket{01}+c^i\ket{10}+d^i\ket{11}$, up to a undetermined overall phase factor,  then yields
\begin{equation}
a^i=a_1^ia_2^i,\quad b^i=a_1^ib_2^i,\quad c^i=a_2^ib_1^i,\quad d^i=b_1^ib_2^i,
\end{equation}
{\em i.e}
\beq
|a^id^i-b^ic^i|=0.
\eeq
It means that $|\bm{\xi}^i_{\rm S}|=1$ and then $|\bm{\xi}^{i\prime\prime}_{\rm S\lambda}|=1$. Of course, $E_{EF}(\rho^i_{\rm S})=E_{EF}(\rho^{i\prime\prime}_{\rm S\lambda})=0$. That is 
\beq
E_{EF}(\rho_{\rm S}^{\prime\prime})=\sum_\lambda q_\lambda\sum_i q_{i\lambda} p_i E_{EF}(\rho^{i\prime\prime}_{\rm S\lambda})=0,
\eeq
where $q_{i\lambda}=\Tr(A_\lambda\otimes B_\lambda)\rho_{\rm S}^i(A_\lambda^\dagger\otimes B_\lambda^\dagger)/q_\lambda$. Because EF is a upper bound of the known measures of entanglement, also one of MRE, we have the conclusion of lemma six. 

\section{Definition of  MRE and Relative Density Matrix}

In the case of pure states, so-called MRE is such a relative entropy of entanglement that its relative density matrix is given definitely. For the mixed states,  we  define MRE by  means of  the physical idea of EF and  information theoretical feature of RE. That is, 

{\bf Definition}.\ For a pure state $\rho^{\rm P}$ and a mixed state $\rho^{\rm M}$, MRE is defined respectively as
\begin{eqnarray}
E_{MRE}(\rho^{\rm P})&=&S(\rho^{\rm P}\|R(\rho^{\rm P}))=E_{EF}(\rho^{\rm P})\label{MREforP}, \\
E_{MRE}(\rho^{\rm M})&=&\min_{\{p_i,\rho^i\}\in{\cal{D}}} S\left(\rho^{\rm M}\|\sum_{i}p_iR(\rho^i)\right)\label{MREforM}\\ 
&=&\min_{\{p_i,\rho^i\}\in{\cal{D}}} S\left(\rho^{\rm M}\|R^{\rm M}\right),
\end{eqnarray}
where $R(\rho^{\rm P})$ is such a relative density matrix corresponding to the pure state $\rho^{\rm P}$ that eq.(\ref{MREforP}) is satisfied and $R(\rho^{\rm P})$ is a disentangled density matrix. Note that the superscript P denotes a pure state and the superscript M denotes a mixed state. 
In eq.(\ref{MREforM}), the minimum is taken over the set ${\cal{D}}$ that includes all the possible decompositions of pure states  $\rho^{\rm M}=\sum_i p_i\rho^i$. While  
\beq
R^{\rm M}=\sum_{i}p_iR(\rho^i)
\eeq
is a relative density matrix for a mixed state in a given pure state decomposition, where each $R(\rho^i)$ is found out by means of eq.(\ref{MREforP}) for the pure state $\rho^i$. In particular, for two qubits, the relative density matrix  can be chosen by the following theorem one. 

{\bf Theorem one}. In the case of the pure state $\rho^{\rm P}$ of two qubits, the relative density matrix of MRE can be taken as
\beq
R_{A\!B}(\rho_{A\!B}^{\rm P})=\sum_{j=1}^{2} q^{(j)}(\rho_{A\!B}^{\rm P}) \bar{\rho}^{(j)}_A(\rho_{A\!B}^{\rm P})\otimes \bar{\rho}^{(j)}_B(\rho_{A\!B}^{\rm P}).
\label{2RDM}
\eeq
The subscript $AB$ denotes bi-party systems, the subscript $A$ and $B$ denote $A$-party and $B$-party respectively. The coefficients $q^{(j)}(\rho_{A\!B}^{\rm P})$ read
\beql
\label{2RDMqa}
q^{(1)}(\rho_{A\!B}^{\rm P})&=&\frac{1-\xi(\rho_{A\!B}^{\rm P})}{2},\\ 
q^{(2)}(\rho_{A\!B}^{\rm P})&=&1-q^{(1)}(\rho_{A\!B}^{\rm P}),
\label{2RDMb}
\eeql
the density matrices $\bar{\rho}^{(j)}_A(\rho_{A\!B}^{\rm P})$ and $\bar{\rho}^{(j)}_B(\rho_{A\!B}^{\rm P})$ respectively for $A$ and $B$ parties are defined by 
\beql
\label{2RDMBa}
\bar{\rho}^{(1)}_A(\rho_{A\!B}^{\rm P})&=&\frac{1}{2}\left[\sigma_0-\bm{\eta}_A(\rho_{A\!B}^{\rm P})\cdot\bm{\sigma}\right],\\
\bar{\rho}^{(1)}_B(\rho_{A\!B}^{\rm P})&=&\frac{1}{2}\left[\sigma_0-\bm{\eta}_B(\rho_{A\!B}^{\rm P})\cdot\bm{\sigma}\right],\\
\bar{\rho}^{(2)}_A(\rho_{A\!B}^{\rm P})&=&\sigma_0-\bar{\rho}^{(1)}_A(\rho_{A\!B}^{\rm P}),\\
\bar{\rho}^{(2)}_B(\rho_{A\!B}^{\rm P})&=&\sigma_0-\bar{\rho}^{(1)}_B(\rho_{A\!B}^{\rm P}),
\label{2RDMd}
\eeql
where $\sigma_0$ is $2\times 2$ identity matrix and $\sigma_k\ (k=1,2,3)$ are usual Pauli Matrices and $\bm{\eta}_A$ and $\bm{\eta}_B$ are defined by
\beql
\bm{\eta}_{A}(\rho_{A\!B}^{\rm P})&=&\frac{\bm{\xi}_{A}(\rho_{A\!B}^{\rm P})}{\xi(\rho_{A\!B}^{\rm P})}\quad ({\xi(\rho_{A\!B}^{\rm P})}\neq 0),\\
\bm{\eta}_{B}(\rho_{A\!B}^{\rm P})&=&\frac{\bm{\xi}_{B}(\rho_{A\!B}^{\rm P})}{\xi(\rho_{A\!B}^{\rm P})}\quad ({\xi(\rho_{A\!B}^{\rm P})}\neq 0),\\
\bm{\eta}_A(\rho_{A\!B}^{\rm P})&=&\pm\bm{\eta}_B(\rho_{A\!B}^{\rm P})=
\{0,0,1\}\quad ({\xi(\rho_{A\!B}^{\rm P})}=0)\label{eta0},
\eeql
where $\bm{\xi}_A$ and $\bm{\xi}_B$ are the polarized vectors of reduced density matrices respectively for $\rho_A$ and $\rho_B$, $\xi=\xi(\rho_{A\!B})$ is their norm. For the maximally entangled states 
\begin{equation}
\ket{\Phi^\pm}=\frac{1}{\sqrt{2}}(\ket{00}\pm\ket{11}),\  \ket{\Psi^\pm}=\frac{1}{\sqrt{2}}(\ket{01}\pm\ket{10}),
\end{equation}
the sign in eq.(\ref{eta0}), is taken as ``$+$" if $\rho_{A\!B}^{\rm P}=\ket{\Phi^\pm}\bra{\Phi^\pm}$, and taken as ``$-$" if $\rho_{A\!B}^{\rm P}=\ket{\Psi^\pm}\bra{\Psi^\pm}$.  That is
\beqa
R(\rho(\Phi^\pm))&=&\frac{1}{2}\left( \frac{1}{2}\left(I+\sigma_3\right)\otimes \frac{1}{2}\left(I+\sigma_3\right)\right)\nonumber\\
& &+\frac{1}{2} \left(\frac{1}{2}\left(I-\sigma_3\right)\otimes \frac{1}{2}\left(I-\sigma_3\right)\right),\\
R(\rho(\Psi^\pm))&=&\frac{1}{2}\left( \frac{1}{2}\left(I+\sigma_3\right)\otimes \frac{1}{2}\left(I-\sigma_3\right)\right)\nonumber\\
& &+\frac{1}{2} \left(\frac{1}{2}\left(I-\sigma_3\right)\otimes \frac{1}{2}\left(I+\sigma_3\right)\right).
\eeqa

We called $\bar{\rho}^{(j)}_A(\rho_{A\!B}^{\rm P}), \bar{\rho}^{(j)}_B(\rho_{A\!B}^{\rm P})$ as the basis of the relative density matrix in a pure state $\rho_{A\!B}^{\rm P}$ respectively for $A$-party and $B$-party. Their meaning can be more clearly seen in MRE for multi-party systems \cite{My1}. It is very easy to verify that the relative entropy calculated in terms of $R(\rho^{\rm P})$ for a pure state is equal to EF and our MRE. 

Now, we explain why we take the relative density matrix defined as above to evaluate MRE. 

In practice, from the knowledge about RE, we understand, if we can find such a relative density matrix $R$ that  $S(\rho||R)\leq S(\rho||\rho^R)$ for arbitrary $\rho^R\in{\cal{R}}$, where  ${\cal{R}}$ consists of all of disentangled states. Thus, by means of lemma one our task is just to find the minimum value of eq.(\ref{CRE}).  Obviously, it is too complicated in terms of standard method, because one has to differentiate $S(\rho||\rho^R)$ to 15 independent parameters in the relative density matrix, gets the equation systems by making these derivatives equal to zero, and then solves this equation system. 

In order to avoid above difficulty,  in the case of pure states, we use a  trick, that is, to chose a particular subset of ${\cal{R}}$ and find the relative density matrix in this subset that not only leads to the minimum value of relative entropy in eq.(\ref{CRE}) but also is equal to the entanglement of formation.  So we can conclude that a correct and suitable relative density matrix for MRE has been found. Actually, if there exists any other relative density matrices $M^R\in{\cal{R}}$ which can result in $S(\rho||M^R)< S(\rho||R)=E_{EF}(\rho)$, it must be contradict with the conclusion that RE is equal to EF for pure states. In other words, only considering a particular subset of ${\cal{R}}$ is enough to find a suitable relative density matrix in MRE. We does not exclude the possibility that there exist other suitable relative density matrices in the set ${\cal{R}}$.  However, they are not needed by us. 

 Based on analysis and argumentation above, we, in eq.(\ref{RE}), choose such a subset $\{\rho^R\}$ of ${\cal{R}}$ that every eigen decomposition state $\rho_\alpha^R$ of $\rho^R$ is purely separable as $\rho_\alpha^R=\rho_{\alpha A}^R\otimes\rho_{\alpha B}^R$. For simplicity, only consider the case with two qubits. Because that the state described by a eigen density matrix is pure, $\rho_A^\alpha$ and $\rho_B^\alpha$ have to be pure. While the $2\times 2$ density matrix can be written as
 \beqa
\rho_{\alpha A}^R=\frac{1}{2}(1+\bm{\eta}_{A}^\alpha\cdot \bm{\sigma}),\\
\rho_{\alpha B}^R=\frac{1}{2}(1+\bm{\eta}_{B}^\alpha\cdot \bm{\sigma}),
\eeqa
Denoting  $\eta_A^\alpha=(1,\bm{\eta}_A^\alpha)=\{\eta_{A \mu}^\alpha,\mu=0,1,2,3\}$ and $\eta_B^\alpha=(1,\bm{\eta}_A^\alpha)=\{\eta_{\mu B}^\alpha,\mu=0,1,2,3\}$, it is easy to obtain that
\beqa 
\label{e38}
\Tr(\rho\rho_\alpha^{\rm R})&=&\frac{1}{16}\sum_{\mu,\nu=0}^3\sum_{\mu^\prime,\nu^\prime=0}^3 a_{\mu\nu} \eta_{A\mu^\prime}^\alpha \eta_{B\nu^\prime}^\alpha\nonumber\\ & &\Tr[(\sigma_\mu\otimes\sigma_\nu)(\sigma_{\mu^\prime}\otimes\sigma_{\nu^\prime})]\nonumber\\
&=&\frac{1}{4} \sum_{\mu,\nu=0}^3 \eta^\alpha_A{}_\mu a_{\mu\nu}\eta^\alpha_B{}_\nu=\omega_\alpha,
\eeqa
where we have used that $(A\otimes B)(C\otimes D)=(AC)\otimes (BD)$ and $\Tr(\sigma_\mu\sigma_\nu)=2\delta_{\mu\nu}$. 

Actually, it is enough for our aim only to find the extreme surface fixing all the eigenvalues of $\rho^{\rm R}$. Suppose first that there is no any zero eigenvalue in $\rho^{\rm R}$ and denote that 
\beql
\label{EE1}
\lambda_1&=&1-x,\\
\lambda_2&=&1-y,\\
\lambda_3&=&1-z,\\
\lambda_4&=&x+y+z-2,
\label{EE2}
\eeql
where $1>x>0,1>y>0,1>z>0$ since each eigenvalue larger than 0 and less than 1. 
Based on lemma one, in terms of eq.(\ref{CRE}) and noting in the case of pure states, the following equation system is obtained 
\beql
\frac{\partial S(\rho\|\rho^{\rm R})}{\partial x}&=&\frac{\omega^1}{1-x}-\frac{\omega^4}{x+y+z-2}=0,\\
\frac{\partial S(\rho\|\rho^{\rm R})}{\partial y}&=&\frac{\omega^2}{1-y}-\frac{\omega^4}{x+y+z-2}=0,\\
\frac{\partial S(\rho\|\rho^{\rm R})}{\partial z}&=&\frac{\omega^3}{1-z}-\frac{\omega^4}{x+y+z-2}=0,
\eeql
it is easy to get
\begin{equation}
\frac{\omega^1}{\omega^2}=\frac{\lambda_1}{\lambda_2},\quad \frac{\omega^2}{\omega^3}=\frac{\lambda_2}{\lambda_3},\quad \frac{\omega^3}{\omega^1}=\frac{\lambda_3}{\lambda_1}.
\end{equation}
We can write their solutions as
\begin{equation}
1-x=\beta\omega^1,\quad 1-y=\beta\omega^2,\quad 1-z=\beta\omega^3.
\end{equation}
Obviously, substituting them back to (\ref{EE1}--\ref{EE2}), we have $\beta=1$. This indicate that 
\beq
\lambda_\alpha=\omega_\alpha.
\eeq
It is easy to verify that this gives out the minimum surface. 
If there are some zero eigenvalues in $\rho^{\rm R}$, we can obtain the same result in the similar way. Therefore, the minimum relative entropy in the surface is
\begin{equation}
S(\rho\|\rho^{\rm R})=-S(\rho)-\sum_{\alpha}\omega^\alpha \log\omega_\alpha.
\end{equation}
From eq.(\ref{e38}), it follows that
\beqa
\omega^\alpha &=&\frac{1}{4}\sum_{\mu,\nu=0}^3 \eta^\alpha_A{}_\mu a_{\mu\nu}\eta^\alpha_B{}_\nu\nonumber\\
&=&\frac{1}{4}(1+\bm{\eta}_A^\alpha\cdot\bm{\xi}_A +\bm{\xi}_B\cdot\bm{\eta}_B^\alpha +\eta_{Ai}^\alpha a_{ij}\eta_{Bj}^\alpha).
\eeqa
 
 Furthermore, in terms of the orthogonal property among the different $\rho_\alpha^R$, when $|\bm{\xi}|\neq 0$,
we can choose $\bm{\eta}_A^1=\bm{\eta}_A^2=-\bm{\eta}_A^3=-\bm{\eta}_A^4$, $\bm{\eta}_B^1=-\bm{\eta}_B^2=\bm{\eta}_B^3=-\bm{\eta}_B^4$, as well as $\bm{\eta}_A^1=\bm{\xi}_A/|\bm{\xi}_A|, \bm{\eta}_B^1=\bm{\xi}_B/|\bm{\xi}_B|$.
From the facts  that their norms are all 1,  $|\bm{\xi}_A|=|\bm{\xi}_B|=|\bm{\xi}|$ in the case of pure states and lemma two, it follows that
\beq
\omega^1=\frac{1}{2}(1+|\bm{\xi}|),\quad \omega^2=\omega^3=0, \quad
\omega^4=\frac{1}{2}(1-|\bm{\xi}|).
\eeq
When $|\bm{\xi}|=0$, we have $\xi_{Ai}=\xi_{Bj}=0, (i,j=1,2,3)$.  Thus
\beq
\omega^\alpha=\frac{1}{4}(1+\eta_{Ai}^\alpha a_{ij}\eta_{Bj}^\alpha).
\eeq
Moreover, from lemma three we know the corresponding quantum states are Bell states with maximum entanglement. Obviously, in this case
\beq
(a_{ij})_{\Phi^+}=\left(\begin{array}{ccc}
1&0&0\\
0&-1&0\\
0&0&1
\end{array}
\right),
\eeq
\beq
(a_{ij})_{\Phi^-}=\left(\begin{array}{ccc}
-1&0&0\\
0&1&0\\
0&0&1
\end{array}
\right),
\eeq
\beq
(a_{ij})_{\Psi^+}=\left(\begin{array}{ccc}
1&0&0\\
0&1&0\\
0&0&-1
\end{array}
\right),
\eeq
\beq
(a_{ij})_{\Psi^-}=\left(\begin{array}{ccc}
-1&0&0\\
0&-1&0\\
0&0&-1
\end{array}
\right).
\eeq
Then, we can choose, for Bell states $\ket{\Phi^\pm}$,  $\bm{\eta}_A^1=\bm{\eta}_A^2=\bm{\eta}_A^3=\bm{\eta}_A^4=(0,0,1)$, $\bm{\eta}_B^1=\bm{\eta}_B^2=-\bm{\eta}_B^3=-\bm{\eta}_B^4=(0,0,1)$; for Bell states $\ket{\Psi^\pm}$, $\bm{\eta}_A^1=\bm{\eta}_A^2=\bm{\eta}_A^3=\bm{\eta}_A^4=(0,0,1)$,  $\bm{\eta}_B^1=\bm{\eta}_B^2=-\bm{\eta}_B^3=-\bm{\eta}_B^4=(0,0,-1)$. It follows that
\beq
\omega^1=\omega^2=\frac{1}{2},\quad \omega^3=\omega^4=0.
\eeq

It is well known that the two non-zero eigenvalues of the reduced density matrix are respectively $\displaystyle \frac{1}{2}(1\pm|\bm{\xi}|)$. Therefore, in the case of pure states, when we take the above relative density matrix defined as the theorem one,  it is obtained immediately 
\beqa
\label{MEFV1}
S(\rho^{\rm P}\|R(\rho^{\rm P}))&=&E_{MRE}
=S(\rho_{\{A,B\}}^{\rm P})\\
&=&E_{EF}(\rho^{\rm P})=E_{RE}(\rho^{\rm P})\label{MEFV2}
\eeqa
in the case of pure states. The subscript $\{A, B\}$ is a compact denotation for $A$ or $B$. In other word, we have found a suitable relative density matrix to calculate MRE, also RE,  for arbitrary pure states of two qubit systems. 

Up to now, we have proved that the theorem one is indeed one solution of the separable relative matrix which leads to the minimum values of relative entropy for a pure state. It is unnecessary to consider more general cases because if there exists other separable relative matrix which leads to the value of relative entropy less than one in eq.(\ref{MEFV1}) or (\ref{MEFV2}), it will broke the well-known theorem that RE for a pure state must be equal to its EF.

In principle, for the systems with more qubits, the relative density matrix $R$ for MRE in a given pure state can be defined and found  by solving equation $S(\rho^{\rm P}\|R)=S(\rho_B^{\rm P})=E_{EF}(\rho^{\rm P})$ based on the fact that EF is a good enough measure of entanglement in this case.  For the case of mixed state, we first find the relative density matrix $R(\rho^i)$, in which $\rho^i$ belong to a pure state decomposition, by solving equation $S(\rho^i\|R(\rho^i))=S(\rho_{B}^{i})=E_{EF}(\rho^i)$. Then, we can write the total relative density matrix for a mixed state as $R^{\rm M}=\sum_{i}p_iR(\rho^i)$. Obviously, for all of pure state decompositions, in terms of this method, one can construct their relative density matrices and calculate the corresponding relative entropies. The last, MRE is obtained by taking the minimum one among these relative entropies. This is just our  algorithm to calculate MRE. 

For two qubit systems, we have successfully obtained the explicit and general expression of the relative density matrix in an arbitrary pure state or a mixed state with any given decomposition. MRE for two qubit systems can be easier calculated because the first step in our algorithm is finished. 
For more than two qubits, we do not give clearly an explicit expression of the relative density matrix for a pure state in this paper.  In fact,  to find relative density matrix needs more computations, but our algorithm still works in principle. This is because that from  $S(\rho^i\|R)=S(\rho_{B}^{i})=E_{EF}(\rho^i)$ to find $R(\rho^i)$ can be done within finite steps for a given pure state in general except for the solution $R(\rho^i)$ does not exist. The exception is impossible because this implies that for the pure state $\rho_i$ RE has no a relative density matrix so as to it correctly measure entanglement, or saying,  it breaks down again the conclusion that for a pure state RE is equal to EF, while the latter always exists in a pure state. 

In addition, it must be emphasized that our method is to calculate MRE but not EF.  Our algorithm of MRE and  Wootter's method for EF can not be replaced each other. In the case of mixed states, MRE is different from EF in general, also from Wootter's EF. In the discussion on Werner state, we will see that EF is linearly depending on the probability of component states, but MRE is logarithmically depending on the probability of component states. In our point of view, perhaps it also seems to be a requirement from quantum physics and information theory, the logarithmic dependence on  the probability of component states is more natural and essential. This is one of main reasons why we take the relative entropy to describe the entanglement in the case of mixed states.  

In above sense, MRE avoids the difficulty of RE to find the relative density matrix from an infinite large set of disentangled states and so improve the computability of RE.  In our paper \cite{My1}, we also have given an explicit expression of the relative density matrix for $n$-party systems (restricted to qubits).  

\section{Examples and Discussions}

It must be emphasized that one of advantages of MRE is to decrease the dependence on pure state decomposition.  For example, the state $M$ has two pure state decompositions
\begin{eqnarray}
M&=&\frac{1}{2}\left(\ket{00}\bra{00}+\ket{11}\bra{11}\right)\\
&=&\frac{1}{2}\left(\ket{\Phi^+}\bra{\Phi^+}+\ket{\Phi^-}\bra{\Phi^-}\right), 
\end{eqnarray}
which respectively correspond to the minimum and maximum decompositions in the calculation of EF. But two decompositions have the same relative density matrices in the calculation of MRE. That is, both of them are the minimum for MRE and can be used to calculate MRE. This means that the minimum decomposition(s) to calculate MRE is (are) not the same as the minimum decomposition(s) to calculate EF in general. The former is easier to be found. 

It is interesting to study the mixture of Bell's state. Its form is
\beqa
B_M&=&b_1\ket{\Phi^+}\bra{\Phi^+}+b_2\ket{\Phi^-}\bra{\Phi^-}\nonumber\\
& &+b_3\ket{\Psi^+}\bra{\Psi^+}+b_4\ket{\Psi^-}\bra{\Psi^-}
\eeqa
Obviously, form Peres's condition \cite{Peres}, it follows that its separable condition is
\beq
b_{\rm max}=\max[b_1,b_2,b_3,b_4]\leq \frac{1}{2}
\eeq
Through calculating, we can find out their MPSD as
\beq
\label{MPSDBM}
B_M=\frac{1}{8}\sum_{x,y,z=0}^1 \ket{\psi(x,y,z)}\bra{\psi(x,y,z)}
\eeq
where the forms of $\ket{\psi(x,y,z)}$ in the following four cases are:

Case one: $b_1$ is the maximum and not less than 1/2
\beqa
\ket{\psi(x,y,z)}&=&\sqrt{b_1}\ket{\Phi^+}+\e^{\I\pi x}\sqrt{b_2}\ket{\Phi^-}\nonumber\\
& &+\e^{\I\pi x}\sqrt{b_3}\ket{\Psi^+}+\I\e^{\I\pi z}\sqrt{b_4}\ket{\Psi^-}
\eeqa

Case two: $b_2$ is the maximum and not less than 1/2
\beqa
\ket{\psi(x,y,z)}&=&\sqrt{b_1}\ket{\Phi^+}+\e^{\I\pi x}\sqrt{b_2}\ket{\Phi^-}\nonumber\\
& &+\I\e^{\I\pi x}\sqrt{b_3}\ket{\Psi^+}+\e^{\I\pi z}\sqrt{b_4}\ket{\Psi^-}
\eeqa

Case there: $b_3$ is the maximum and not less than 1/2
\beqa
\ket{\psi(x,y,z)}&=&\sqrt{b_1}\ket{\Phi^+}+\I\e^{\I\pi x}\sqrt{b_2}\ket{\Phi^-}\nonumber\\
& &+\e^{\I\pi x}\sqrt{b_3}\ket{\Psi^+}+\e^{\I\pi z}\sqrt{b_4}\ket{\Psi^-}
\eeqa

Case four: $b_4$ is the maximum and not less than 1/2
\beqa
\ket{\psi(x,y,z)}&=&\sqrt{b_1}\ket{\Phi^+}+\I\e^{\I\pi x}\sqrt{b_2}\ket{\Phi^-}\nonumber\\
& &+\I\e^{\I\pi x}\sqrt{b_3}\ket{\Psi^+}+\I\e^{\I\pi z}\sqrt{b_4}\ket{\Psi^-}
\eeqa

In terms of our theorem one, we can find the relative density matrix for the pure state $\ket{\psi(x,y,z)}$ and construct the total relative density matrix for MPSD of the mixture of Bell's states (\ref{MPSDBM}). Then, we can construct the total relative density matrix for the four case
\end{multicols}
\beq
R_T(B_W|b_1=b_{\max})=\frac{1}{4}\left(
\begin{array}{cccc}
\displaystyle 1+\frac{b_2}{b_2+b_3+b_4}&0&0&\displaystyle1-\frac{b_2}{b_2+b_3+b_4}\\[10pt]
\displaystyle 0&\displaystyle1-\frac{b_2}{b_2+b_3+b_4}&\displaystyle \frac{b_3-b_4}{b_2+b_3+b_4}&0\\[10pt]
\displaystyle 0&\displaystyle\frac{b_3-b_4}{b_2+b_3+b_4}&\displaystyle 1-\frac{b_2}{b_2+b_3+b_4}&0\\[10pt]
\displaystyle1-\frac{b_2}{b_2+b_3+b_4}&0&0&\displaystyle 1+\frac{b_2}{b_2+b_3+b_4}
 \end{array}
 \right)
 \eeq
\beq
R_T(B_W|b_2=b_{\max})=\frac{1}{4}\left(
\begin{array}{cccc}
\displaystyle 1+\frac{b_1}{b_1+b_3+b_4}&0&0&\displaystyle -1+\frac{b_1}{b_1+b_3+b_4}\\[10pt]
\displaystyle 0&\displaystyle1-\frac{b_1}{b_1+b_3+b_4}&\displaystyle \frac{b_3-b_4}{b_1+b_3+b_4}&0\\[10pt]
\displaystyle 0&\displaystyle\frac{b_3-b_4}{b_1+b_3+b_4}&\displaystyle 1-\frac{b_1}{b_1+b_3+b_4}&0\\[10pt]
\displaystyle -1+\frac{b_1}{b_1+b_3+b_4}&0&0&\displaystyle 1+\frac{b_1}{b_1+b_3+b_4}
 \end{array}
 \right)
 \eeq
\beq
R_T(B_W|b_3=b_{\max})=\frac{1}{4}\left(
\begin{array}{cccc}
\displaystyle 1-\frac{b_4}{b_1+b_2+b_4}&0&0&\displaystyle \frac{b_1-b_2}{b_1+b_2+b_4}\\[10pt]
\displaystyle 0&\displaystyle1+\frac{b_4}{b_1+b_2+b_4}&\displaystyle 1-\frac{b_4}{b_1+b_2+b_4}&0\\[10pt]
\displaystyle 0&\displaystyle1-\frac{b_4}{b_1+b_2+b_4}&\displaystyle 1+\frac{b_4}{b_1+b_2+b_4}&0\\[10pt]
\displaystyle \frac{b_1-b_2}{b_1+b_2+b_4}&0&0&\displaystyle 1-\frac{b_4}{b_1+b_2+b_4}
 \end{array}
 \right)
 \eeq
 \beq
R_T(B_W|b_4=b_{\max})=\frac{1}{4}\left(
\begin{array}{cccc}
\displaystyle 1-\frac{b_3}{b_1+b_2+b_3}&0&0&\displaystyle \frac{b_1-b_2}{b_1+b_2+b_3}\\[10pt]
\displaystyle 0&\displaystyle1+\frac{b_3}{b_1+b_2+b_3}&\displaystyle -1+\frac{b_3}{b_1+b_2+b_3}&0\\[10pt]
\displaystyle 0&\displaystyle -1+\frac{b_3}{b_1+b_2+b_3}&\displaystyle 1+\frac{b_3}{b_1+b_2+b_3}&0\\[10pt]
\displaystyle \frac{b_1-b_2}{b_1+b_2+b_3}&0&0&\displaystyle 1-\frac{b_3}{b_1+b_2+b_3}
 \end{array}
 \right)
 \eeq
\begin{multicols}{2}
Then we can calculate out MRE of the mixture of Bell's state in all of four cases
\beqa
E_{\rm MRE}(B_M)&=&\log 2+b_{\max}\log b_{\max}\nonumber\\
& &+(1-b_{\max})\log(1- b_{\max})
\eeqa
when $b_{\max}\geq 1/2$. Compare with EF of the mixture of  Bell's states
\beqa
E_{\rm EF}(B_M)&=&-\frac{1-2\sqrt{b_{\max}(1-b_{\max})}}{2}\nonumber\\
& &\log\left[\frac{1-2\sqrt{b_{\max}(1-b_{\max})}}{2}\right]\nonumber\\
& &-\frac{1+2\sqrt{b_{\max}(1-b_{\max})}}{2}\nonumber\\
& &\log \left[\frac{1+2\sqrt{b_{\max}(1-b_{\max})}}{2}\right]
\eeqa
We can see the similar feature between them, that is, both of them are only dependent on the maximum eigenvalue $b_{\max}$. Obviously, MRE's behavior is   acceptable because of its real logarithmic dependence on the probability of component states in MPSD. Moreover, $E_{\rm MRE}(B_M) \leq E_{\rm EF}(B_M)$. This is just an expected property. However, it is necessary to study carefully the purification and concentration of entanglement so as to account for why it is so and which is better. Since there exist some difficulties in EF for purification of the mixed state, it is significant and interesting to research MRE. In fact, comparing with EF, MRE has not lost any physical information and content in measuring entanglement of the pure states. While in the case of mixed states, MRE appears to play a hopeful role. 

In special, for Werner's state \cite{Werner} 
\begin{eqnarray}
W&=&F\ket{\Psi^-}\bra{\Psi^-}+\frac{1-F}{3}(\ket{\Psi^+}\bra{\Psi^+}\nonumber\\ 
& &+\ket{\Phi^+}\bra{\Phi^+}+\ket{\Phi^-}\bra{\Phi^-}).
\end{eqnarray}
When $F\geq 1/2$, we can find the total relative matrix
\beq
R_T(W)=\frac{1}{6}\left(
\begin{array}{cccc}
\displaystyle1&0&0&0\\
\displaystyle 0&\displaystyle 2&\displaystyle -1&0\\
\displaystyle 0&\displaystyle -1&\displaystyle 2&0\\
\displaystyle 0&0&0&\displaystyle 1
 \end{array}
 \right)
 \eeq
Therefore,  the modified relative entropy of entanglement of Werner's state is found to be
\beq
E_{MRE}(W)=\log 2+F\log F+(1-F)\log(1-F)
\eeq
Note that the entanglement of formation of Werner's state is
\beqa
E_{EF}(W)&=&-\frac{1-2\sqrt{F(1-F)}}{2}\nonumber\\
& &\log\left[\frac{1-2\sqrt{F(1-F)}}{2}\right]\nonumber\\ 
& &-\frac{1+2\sqrt{F(1-F)}}{2}\nonumber\\
& &\log\left[\frac{1+2\sqrt{F(1-F)}}{2}\right]
\eeqa
when $F\geq 1/2$. They can be displayed as the following figure
\begin{figure}
\begin{center}
\epsfxsize=2.7in
\epsfysize=2.3in
\epsfbox{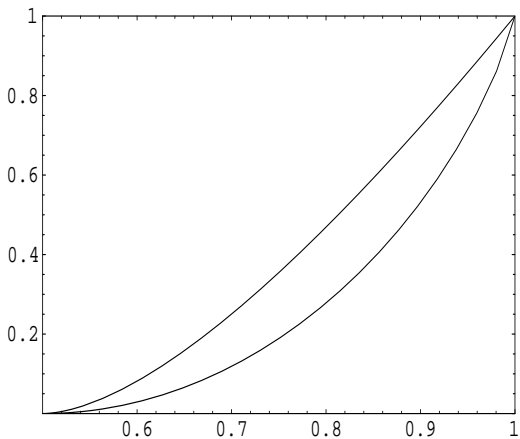}
\vskip 0.1in
\caption{The above curve is $E_{EF}(W)$ varying curve with $F$, The underside curve is $E_{MRE}(W)$ varying curve with $F$.}
\label{fig1}
\end{center}
\end{figure}
\vskip -0.15in
This figure also can display the varying of MRE and EF with $b_{\max}$ from $1/2$ to $1$ for the mixture of Bell's states. 

It is also interesting to consider the departure state from Bell's states. Set they are
\beq
B^D_i=G \ket{\Psi^-}\bra{\Psi^-}+(1-G) \ket{i}\bra{i}
\eeq
where $\ket{i}$ takes over $\ket{00},\ket{01},\ket{10},\ket{11}$ respective with $i=1,2,3,4$. We can find out their pure state decompositions
\beq
\label{BDPSD}
B^D_i=\frac{1}{2}\ket{\psi_i(+)}\bra{\psi_i(+)}+\frac{1}{2}\ket{\psi_i(-)}\bra{\psi_i(-)}
\eeq
where
\beq
\ket{\psi_i(\pm)}=\sqrt{\lambda_{i1}}\ket{i}\pm\sqrt{\lambda_{i2}}\ket{\Psi^-}
\eeq
while $\lambda_{i1}$ and $\lambda_{i2}$  are respectively  the eigenvalues of $B^D_i$ with the eigenvectors $\ket{i}$ and $\ket{\Psi^-}$. 

For $B^D_1$ and $B^D_4$, they are the minimum pure state decomposition for EF, and then their entanglement of formation are just
\beqa
E_{\rm EF}(B^D_{1,4})&=&-\frac{1-\sqrt{1-G^2}}{2}\log \left( \frac{1-\sqrt{1-G^2}}{2}\right)\nonumber\\
& &-\frac{1+\sqrt{1-G^2}}{2}\log \left(\frac{1+\sqrt{1-G^2}}{2}\right)
\label{BDEF14}
\eeqa
But for $B^D_2$ and $B^D_3$, the decomposition (\ref{BDPSD}) for EF may not to be  suitable.  This is because, based on this decomposition, the statistic average of the reduced entropy of their component states defined by
\beqa
\bar{S}(B^D_i)&=&\frac{1}{2}H\left(\frac{1+\xi_i(+)}{2}\right)+\frac{1}{2}H\left(\frac{1+\xi_i(-)}{2}\right)\\
\xi_i(\pm)&=&\frac{1}{1-2G+2G^2}(G(2-5G+4G^2\nonumber\\
& &\pm 2(1-G)\sqrt{2G(1-G)(1-2G+2G^2)}
\eeqa
has unexpected behaviors varying with $G$ (see Fig.3). In above equation $H(x)=-x\log x-(1-x)\log(1-x)$ is binary entropy function and $i=2,3$. Although the two examples are simple enough, it seems not to be easy to find out their minimum pure state decompositions. 

Fortunately, we have found that the decomposition (\ref{BDPSD}) are correct MPSDs for our modified relative entropy of entanglement. This again implies that MRE decreases the sensitivity on the pure state decompositions.  From our theorems in the above section, we can construct the total relative density matrices 
\beq
R_{T}(B^D_1)=\frac{1}{2(1+G)}\left(
\begin{array}{cccc}
2-G^2&0&0&-G\\
0&G&-G&0\\
0&-G&G&0\\
-G&0&0&G^2
\end{array}\right)
\eeq
\beq
R_{T}(B^D_2)=\left(
\begin{array}{cccc}
0&0&0&0\\
0&1-G/2&0&0\\
0&0&G/2&0\\
0&0&0&0
\end{array}\right)
\eeq
\beq
R_{T}(B^D_3)=\left(
\begin{array}{cccc}
0&0&0&0\\
0&G/2&0&0\\
0&0&1-G/2&0\\
0&0&0&0
\end{array}\right)
\eeq
\beq
R_{T}(B^D_4)=\frac{1}{2(1+G)}\left(
\begin{array}{cccc}
G^2&0&0&-G\\
0&G&-G&0\\
0&-G&G&0\\
-G&0&0&2-G^2
\end{array}\right)
\eeq
Therefore, we can calculate out their MRE
\beqa
E_{\rm MRE}(B^D_{1,4})&=&\frac{1-G}{2}\log\left(1-G\right)+\frac{1+G}{2}\log\left(1+G)\right)\nonumber\\
& &\frac{(1-G)(1-G^2)}{2\sqrt{1-G^2+G^4}} \log\left(\frac{1-\sqrt{1-G^2+G^4}}{1+\sqrt{1-G^2+G^4}}\right)\nonumber\\
& &-(1-G)\log\left(\frac{G}{2}\right)
\eeqa
\beqa
E_{\rm MRE}(B^D_{2,3}) &=&\frac{1}{2}\left[(1-\sqrt{1-2G(1-G)})  \right. 
\nonumber\\
& &\log(1-\sqrt{1-2G(1-G)})\nonumber\\
& &+(1+\sqrt{1-2G(1-G)})\nonumber\\
& &\log(1+\sqrt{1-2G(1-G)})\nonumber\\
& &\left. -G\log G-(2-G)\log\left(2-G)\right)\right]
\eeqa
Comparison EF with MRE for $B^D_1$ and $B^D_4$  can be displayed as the following figure
\begin{figure}
\begin{center}
\epsfxsize=2.7in
\epsfysize=2.3in
\epsfbox{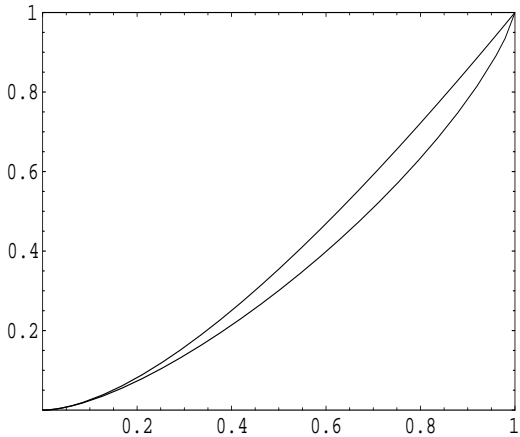}
\vskip 0.1in
\caption{The above curve is $E_{EF}(B^D_{1,4})$ varying curve with $G$, The underside curve is $E_{MRE}(B^D_{1,4})$ varying curve with $G$.}
\label{fig2}
\end{center}
\end{figure}
\vskip -0.15in

A way to improve the unexpected behavior of $\bar{S}(B^D_2)$ and $\bar{S}(B^D_3)$ is to calculate the (modified) EF by means of Wootters' method \cite{Wootters}. In order to compare MRE and Wootters' EF as well as $\bar{S}$, we draw out the following figure
\begin{figure}
\begin{center}
\epsfxsize=2.7in
\epsfysize=2.3in
\epsfbox{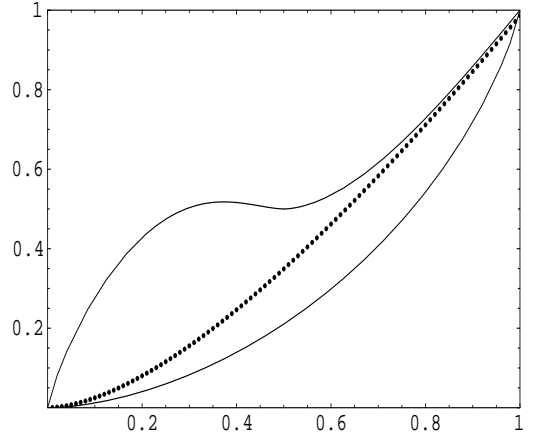}
\vskip 0.1in
\caption{The above curve is $\bar{S}(B^D_{2,3})$ varying curve with $G$, The underside curve is $E_{MRE}(B^D_{2,3})$ varying curve with $G$. The dot curve is Wootters' EF for $B^D_{2,3}$}
\label{fig3}
\end{center}
\end{figure}
\vskip -0.15in

It must be emphasized an important fact that $E_{MRE}(B^D_{1,4})$ is different from $E_{MRE}(B^D_{2,3})$, which can be displayed by
\begin{figure}
\begin{center}
\epsfxsize=2.7in
\epsfysize=2.3in
\epsfbox{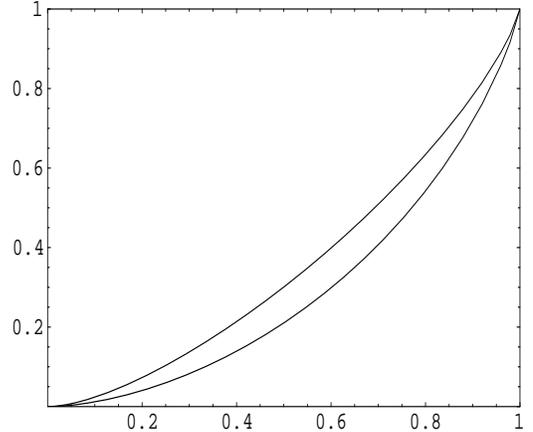}
\vskip 0.1in
\caption{The above curve is $E_{MRE}(B^D_{1,4})$ varying curve with $G$, The underside curve is $E_{MRE}(B^D_{2,3})$ varying curve with $G$. }
\label{fig4}
\end{center}
\end{figure}
\vskip -0.15in
It implies that MRE can reveal the particular difference among them. However, Wootters' EF for them are the same and can be fitted by the eq.(\ref{BDEF14}).  In fact, one can verify that many similar departure states from Bell's state have the same Wootters' EF. In our point of view, there should exist the particular difference among them. It is helpful for a deep and exact description of entanglement. Of course, a possible examination whether MRE is needed and useful can be done by means of identifying such difference from the purification. We will discuss it in our another paper. 

In the following sections, we will further study the properties of MRE. 

\section{Important properties of MRE}

First, we can obtain:

{\bf Theorem two}\ \ Modified relative entropy of entanglement (MRE) is a lower bound of entanglement of formation (EF):
\begin{equation}
E_{MRE}(\rho)\leq E_{EF}(\rho).
\end{equation}
When $\rho$ is a pure state, the equality is valid. 

It is easy to prove it in terms of the joint convexity of the relative entropy
\begin{equation}
S(\sum_i p_i\rho^i\|\sum_i p_i R(\rho^i))\leq \sum_i p_iS(\rho^i\|R(\rho^i))
\end{equation}
and the definition of $E_{EF}$ in eq.(\ref{EF}). Obviously for a pure state, MRE is equal to RE and EF. 

Then, we can see:

{\bf Theorem Three}:\ Modified relative entropy of entanglement (MRE) is a upper bound of relative entropy of entanglement (RE), also one of distillable  entanglement (DE) :
\beq
E_{MRE}(\rho)\geq E_{RE}(\rho)\geq E_{DE}(\rho).
\eeq
When $\rho$ is a pure state, the equality is valid. 

The proof of theorem three is very easy. Because we take a particular disentangled state to calculate MRE, it must be not less than RE. It is also well known that RE is not less than DE and then MRE is not less than DE.  However, we can not prove strictly that the given relative density matrix in MRE is just a disentangled state to give out RE because the set disentangled states is so large that we can not express all of them.  This difficulty is, in fact, from the undetermined feature of RE in computation.   

From theorem two and three, DE$\leq$RE$\leq$MRE$\leq$EF. Noting the fact that both RE and MRE are defined by the relative entropy, we think that MRE is able to inherit most of important physical features of RE if these features of RE are given and proved in terms of the fact stated above as well as some mathematical skills \cite{Vedral1,Vedral2}. 
In fact, we have seen that MRE is a function of the norm of  polarization vectors of the reduced density matrices of the decomposition density matrices for two qubit systems. Thus, both EF and MRE  belong to a kind of the generalized measures of entanglement proposed by \cite{My2}, and the generalized measures of entanglement with the known properties as a good measure are proved there.  In this paper, the behavior of MRE under local general measurement (LGM) and classical communication (CC) can be proved by using of the similar methods  at least for two qubit systems. 

{\bf Theorem Four}\ Any $LGM+CC$ quantum operation does not increase MRE in the case of pure state. 

{\bf Proof.}\ Please note the following facts: (1) We have proved $E_{MRE}(\rho)=E_{EF}(\rho)$ in the case of pure state (theorem one and theorem two); (2) They are both monotone decreasing functions of the norms of the polarized vectors of reduced density matrices for pure states (lemma three); (3) In general,  a pure state will transform to a mixed state under $LGM+CC$.  Obviously, there is a relationship between $\bm{\xi}_{\lambda}^{\prime\prime}{}^2$ and $\bm{\xi}^2$ as following 
\begin{equation}
\bm{\xi}_{\lambda}^{\prime\prime\;2}=\bm{\xi}^2+4|ad-bc|^2\left(1-\frac{\det(A_\lambda^\dagger A_\lambda B_\lambda^\dagger B_\lambda)}{q_\lambda^2}\right).
\end{equation}
Here lemma four has been used. Thus, our aim is convert to prove $\bm{\xi}_{\lambda}^{\prime\prime\;2}\geq\bm{\xi}^2$, that is, $q_\lambda^2\geq\det(A_\lambda^\dagger A_\lambda B_\lambda^\dagger B_\lambda)$. In fact, we can rewrite 
\beq
q_\lambda^2=\bra{\psi}(A^\dagger_\lambda A_\lambda) \otimes (B_\lambda^\dagger B_\lambda)\ket{\psi},
\eeq
where $\rho_{A\!B}=\ket{\psi}\bra{\psi}$. If $(A^\dagger_\lambda A_\lambda) \otimes (B_\lambda^\dagger B_\lambda)$ has any zero eigenvalue, then
\beq
\det(A_\lambda A_\lambda^\dagger  B_\lambda B_\lambda^\dagger )=\sqrt{\det[(A^\dagger_\lambda A_\lambda) \otimes (B_\lambda^\dagger B_\lambda)]}=0.
\eeq
So, we only need to consider the case without zero eigenvalues. Set
\beq
A^\dagger_\lambda A_\lambda =\sum_{\mu=0}^3 c_{\lambda  A}^\mu\sigma_\mu;\quad, B^\dagger_\lambda B_\lambda =\sum_{\mu=0}^3 c_{\lambda B}^\mu\sigma_\mu.
\eeq
Then
\beqa
\det(A_\lambda^\dagger A_\lambda B_\lambda^\dagger  B_\lambda )&=&(c_{\lambda A}^{0\ 2}-\bm{c}_{\lambda A}^2)(c_{\lambda B}^{0\ 2}-\bm{c}_{\lambda B}^2),\\
q_\lambda&=&\sum_{\mu,\nu=0}^3 c_{\lambda A}^\mu a_{\mu\nu}c_{\lambda B}^\nu,
\eeqa
where $a_{\mu\nu}$ is expanding coefficients in eq.(\ref{2qrho}). Because for any states $\ket{\psi}$, $\bra{\psi}A_\lambda^\dagger A_\lambda\ket{\psi}=\|A\ket{\psi}\|^2\geq 0 $ and $\bra{\psi}B_\lambda^\dagger B_\lambda\ket{\psi}=\|B\ket{\psi}\|^2\geq 0 $, we have $A_\lambda^\dagger A_\lambda$ and $B_\lambda^\dagger B_\lambda$ are positive,  then $\Tr (A_\lambda^\dagger A_\lambda)$ and $\Tr(B_\lambda^\dagger B_\lambda)$ are positive, that is,  $c_{\lambda A}^0\geq 0,c_{ \lambda B}^0\geq 0$. Again from 
\beq
\sum_\lambda\Tr[(A^\dagger_\lambda A_\lambda) \otimes (B_\lambda^\dagger B_\lambda)]=4=4\sum_{\lambda}c_{\lambda A}^0c_{\lambda B}^0,
\eeq
it follows that $c_{\lambda A}^0c_{\lambda B}^0\leq 1$.  
Without loss of generality, we can take
\beq
\label{eq107}
c_{\lambda A}^0\leq 1,\quad c_{\lambda B}^0\leq 1,
\eeq
because it is always allowed by multiplying a suitable factor to $A_\lambda$ and dividing $B_\lambda$ by the same factor. From  the facts that $A_\lambda^\dagger A_\lambda\geq 0$ and $B_\lambda^\dagger B_\lambda$ are positive and eq.(\ref{eq107}), it follows that
\beq
\bm{c}_{\lambda A}^2\leq 1,\quad \bm{c}_{\lambda B}^2\leq 1.
\eeq

We can divide $A_\lambda\otimes B_\lambda$ into $(A_\lambda\otimes I)(I\otimes B_\lambda)$. Thus, for the first step transformation
\beq
\bm{\xi}_{\lambda}^{\prime\;2}=\bm{\xi}^2+4|ad-bc|^2\left(1-\frac{\det( B_\lambda^\dagger B_\lambda)}{q^2_{\lambda B}}\right),
\eeq
where
\beqa
q_{\lambda B}&=&\Tr [(I \otimes B_\lambda)\rho_{A\!B}(I\otimes B^\dagger)]\nonumber\\
&=&\sum_{\nu=0}^3 a_{0\nu} c_{\lambda B}^\nu=c_{\lambda B}^0+\bm{\xi}_B\cdot\bm{c}_{\lambda B}.
\eeqa
 Since $\det(B^\dagger_\lambda B_\lambda)=c_{\lambda B}^{0\ 2}-\bm{c}_{\lambda B}^2$ and $c_{\lambda B}^0\geq |\bm{c}_{\lambda B}|$,  we have
\beqa
& & q^2_\lambda-\det(B_\lambda^\dagger B_\lambda)\nonumber\\
&=&2c_{\lambda B}^0\bm{\xi}_B\cdot\bm{c}_{\lambda B}+(\bm{\xi}_B\cdot\bm{c}_{\lambda B})^2+\bm{c}_{\lambda B}^2\nonumber\\
&\geq &\bm{c}_{\lambda B}^2+2|\bm{c}_{\lambda B}|\bm{\xi}_B\cdot\bm{c}_{\lambda B}+(\bm{\xi}_B\cdot\bm{c}_{\lambda B})^2\nonumber\\
&=&(|\bm{c}_{\lambda B}| +\bm{\xi}_B\cdot\bm{c}_{\lambda B})^2\geq 0.
\eeqa
It means that
\beq
\bm{\xi}_{\lambda}^{\prime\;2}\geq\bm{\xi}^2.
\eeq
For the second step transformation, we have
\beq
\bm{\xi}_{\lambda}^{\prime\prime\;2}=\bm{\xi}^{\prime\;2}+4|\bar{a}^\prime \bar{d}{}^\prime-\bar{b}^\prime \bar{c}^\prime|^2\left(1-\frac{\det( A_\lambda^\dagger A_\lambda)}{q^2_{\lambda A}}\right),
\eeq
where
\beqa
q_{\lambda A}&=&\Tr [(A_\lambda \otimes I)\rho_{A\!B}^{\lambda\prime}(A_\lambda^\dagger\otimes B)]\nonumber\\
&=&\sum_{\mu=0}^3 \bar{a}^\prime_{\mu 0} c_{\lambda A}^\mu=c_{\lambda A}^0+\bm{\xi}_{A \lambda }^\prime\cdot\bm{c}_{\lambda A}
\eeqa
and
\beqa
\rho^{\lambda\prime}_{A\!B}&=&\ket{\psi_\lambda^\prime}\bra{\psi_\lambda^\prime}\nonumber\\
&=&(I\otimes B_\lambda)\rho_{A\!B}(I\otimes B_\lambda^\dagger)/q_{\lambda B}\nonumber\\
&=&\frac{1}{4}\sum_{\mu,\nu=0}^3\bar{a}^\prime_{\mu\nu}\sigma_\mu\otimes\sigma_\nu,\\
\ket{\psi_\lambda^\prime}&=&\frac{1}{\sqrt{q_{\lambda B}}}(I\otimes B_\lambda)\ket{\psi}\nonumber\\
&=&\bar{a}^\prime\ket{00}+\bar{b}^\prime\ket{01}+\bar{c}^\prime\ket{10}+\bar{d}{}^\prime\ket{11}\\
(\bm{\xi}^\prime_{\lambda A})^i&=&\bar{a}^\prime_{i0}.
\eeqa
Likewise, we can prove
\beq
\bm{\xi}_{\lambda}^{\prime\prime\;2}\geq\bm{\xi}^{\prime\;2}\geq\bm{\xi}^2.
\eeq
Therefore
\beq
E_{MRE}(\rho_{A\!B}^{\prime\prime\;\lambda})\leq E_{MRE}(\rho_{A\!B}).
\eeq
Of course
\beq
E_{MRE}(\rho_{A\!B}^{\prime\prime})\leq\sum_{\lambda} q_\lambda E_{MRE}(\rho_{A\!B}^{\prime\prime\;\lambda})\leq  E_{MRE}(\rho_{A\!B}).
\eeq
The proof of theorem four is finished. 

{\bf Theorem Five}\ Suppose under $LGM+CC$ quantum operation, $\rho^{\rm M}\longrightarrow \displaystyle \sum_\lambda (A_\lambda\otimes B_\lambda)\rho^{\rm M}(A_\lambda^\dagger\otimes B_\lambda^\dagger)$. When $(A^\dagger_\lambda A_\lambda)\otimes (B^\dagger_\lambda B_\lambda)$ is proportional  to an identity matrix, this $LGM+CC$ quantum operation does not increase MRE in the case of mixed states. 

{\bf Proof}: Now, we consider the case of mixed states. Without loss of generality, we assume we have had a minimum decomposition $\rho^{\rm M}=\sum_i p_i\rho^i$, where each $\rho^i$ is a pure state. Moreover, the relative density matrix $R^{\rm M}$ of MRE is constructed in terms of this decomposition. Obviously
\beq
\label{dli}
\rho^{\rm M}{}^{\prime\prime}=\sum_\lambda q_\lambda \rho^{\rm M}_\lambda{}^{\prime\prime}=\sum_\lambda q_\lambda \sum_i p_i q_{i\lambda}\rho_\lambda^{i\prime\prime},
\eeq
where
\beqa
\rho^{\rm M}_\lambda{}^{\prime\prime}&=&\frac{1}{q_\lambda}(A_\lambda\otimes B_\lambda)\rho^{\rm M}(A_\lambda^\dagger\otimes B_\lambda^\dagger),\\
\rho_\lambda^{i\prime\prime}&=&\frac{1}{q_\lambda q_{i\lambda}}(A_\lambda\otimes B_\lambda)\rho^i(A_\lambda^\dagger\otimes B_\lambda^\dagger),\\
q_\lambda&=&\Tr[(A_\lambda\otimes B_\lambda)\rho^{\rm M}(A_\lambda^\dagger\otimes B_\lambda^\dagger)],\\
q_{i\lambda}&=&\frac{1}{q_\lambda}\Tr[(A_\lambda\otimes B_\lambda)\rho^i(A_\lambda^\dagger\otimes B_\lambda^\dagger)].
\eeqa

Noting that 
\beq
\rho^i=\frac{1}{4}\sum_{\mu\nu} a_{\mu\nu}^i \sigma_\mu\otimes\sigma_\nu,
\eeq
we have
\beq
\rho^{i\prime\prime}_\lambda=\frac{1}{4 q_\lambda q_{i\lambda}}\sum_{\mu,\nu=0}^3 a_{\mu\nu}^i (A_\lambda \sigma_\mu A^\dagger_\lambda)\otimes(B_\lambda \sigma_\nu B^\dagger_\lambda).
\eeq
From the precondition  $(A^\dagger_\lambda A_\lambda)\otimes (B^\dagger_\lambda B_\lambda) \propto I_{4\times 4}$ ,  it follows that $A^\dagger_\lambda A_\lambda\propto I_{2\times 2}$ and $B^\dagger_\lambda B_\lambda\propto I_{2\times 2}$.  Otherwise it will contradict with this precondition. Without loss of generality, suppose $A^\dagger_\lambda A_\lambda=\alpha_\lambda I_{2\times 2}$ and $B^\dagger_\lambda B_\lambda=\beta_\lambda I_{2\times 2}$.Thus
\beqa
\rho^{i\prime\prime}_{\lambda A}=\frac{1}{2 \alpha_\lambda} \sum_{\mu=0}^3 a_{\mu 0}^i (A_\lambda \sigma_\mu A^\dagger_\lambda),\\
\rho^{i\prime\prime}_{\lambda B}=\frac{1}{2 \beta_\lambda}\sum_{\nu=0}^3 a_{0\nu}^i (B_\lambda \sigma_\nu B^\dagger_\lambda),
\eeqa
where we have used the facts that
\beqa
q_\lambda=\Tr[(A_\lambda\otimes B_\lambda)\rho^{\rm M}(A_\lambda^\dagger\otimes B_\lambda^\dagger)]=\alpha_\lambda \beta_\lambda,\\
q_{i\lambda}=\frac{1}{q_\lambda}\Tr[(A_\lambda\otimes B_\lambda)\rho^i(A_\lambda^\dagger\otimes B_\lambda^\dagger)]=1.
\eeqa
From definition of polarized vector, it follows that
\beqa
\bm{\xi}_{\lambda A}^{i\prime\prime}=\frac{1}{2 \alpha_\lambda} \sum_{k=1}^3 a_{k 0}^i\Tr(A_\lambda \sigma_k A^\dagger_\lambda \bm{\sigma}),\\
\bm{\xi}_{\lambda B}^{i\prime\prime}=\frac{1}{2 \beta_\lambda} \sum_{k=1}^3 a_{0 k}^i\Tr(B_\lambda \sigma_k B^\dagger_\lambda \bm{\sigma}).
\eeqa
By using of theorem one, we have
\beqa
R^{\prime\prime}_\lambda(\rho^{\rm M})&=&\frac{1}{q_\lambda}(A_\lambda\otimes B_\lambda)R(\rho^{\rm M})(A_\lambda^\dagger\otimes B_\lambda^\dagger)\\
&=&\sum_{j=1}^{2} q^{(j)}(\rho^i) \bar{\rho}^{(j)\prime\prime}_{A\lambda}(\rho^i)\otimes \bar{\rho}^{(j)\prime\prime}_{B\lambda}(\rho_{A\!B}^{\rm P}),
\eeqa
where 
\beqa
\bar{\rho}^{(j)\prime\prime}_{A\lambda}(\rho^i)=\frac{1}{\alpha_\lambda}(A_\lambda \bar{\rho}^{(i)}A_\lambda^\dagger),\\
\bar{\rho}^{(j)\prime\prime}_{B\lambda}(\rho^i)=\frac{1}{\beta_\lambda}(B_\lambda \bar{\rho}^{(i)}B_\lambda^\dagger).
\eeqa
It is easy to obtain that
\beql
\label{etappA}
\bm{\eta}_{ A\lambda}^{\prime\prime}(\rho^i)&=& 2\Tr\left[\bar{\rho}^{(j)\prime\prime}_{A\lambda}(\rho^i)\bm{\sigma}\right]\nonumber\\
&=&\frac{1}{\alpha_\lambda\xi(\rho^i)}\Tr[(\bm{\xi}_{i A}\cdot A_\lambda \bm{\sigma} A_\lambda^\dagger)\bm{\sigma}]\nonumber\\
&=&\frac{\bm{\xi}_{\lambda A}^{\prime\prime}(\rho^i)}{\xi(\rho^{i\prime\prime}_\lambda)}
=\bm{\eta}_{ A\lambda}(\rho^{i\prime\prime}),\\
\label{etappB}
\bm{\eta}_{ B\lambda}^{\prime\prime}(\rho^i)&=& 2\Tr\left[\bar{\rho}^{(j)\prime\prime}_{B\lambda}(\rho^i)\bm{\sigma}\right]\nonumber\\
&=&\frac{1}{\beta_\lambda\xi(\rho^i)}\Tr[(\bm{\xi}_{i B}\cdot B_\lambda \bm{\sigma} B_\lambda^\dagger)\bm{\sigma}]\nonumber\\
&=&\frac{\bm{\xi}_{\lambda B}^{\prime\prime}(\rho^i)}{\xi(\rho^{i\prime\prime}_\lambda)}=\bm{\eta}_{ B\lambda}(\rho^{i\prime\prime}),
\eeql
when $\xi(\rho^i)\neq 0$. Here, we have used lemma five, that is, $|\bm{\xi}^i|=|\bm{\xi}^{i\prime\prime}_\lambda|$. If $\xi(\rho^i)=0$, that is $\rho^i$ is a maximum entangled state, we have to introduce an infinite small shift for coefficients of states 
\beqa
\label{bellphie}
\ket{\Phi^\pm}_\epsilon=\sqrt{\frac{1-\epsilon}{2}}\ket{00}\pm\sqrt{\frac{1+\epsilon}{2}}\ket{11},\\
\label{bellpsie}
\ket{\Psi^\pm}_\epsilon=\sqrt{\frac{1-\epsilon}{2}}\ket{01}\pm\sqrt{\frac{1+\epsilon}{2}}\ket{10}.
\eeqa
Obviously $(\bm{\xi}^i)^2=\epsilon^2\neq 0$. Then, replacing the maximum states $\rho^i$ by the shifted state $\rho_\epsilon^i$, which consists of $\ket{\Phi^\pm}_\epsilon$ or $\ket{\Psi^\pm}_\epsilon$, we can prove the same conclusion as eqs. (\ref{etappA}) and (\ref{etappB}). So  we immediately  have  the relation
\beqa
\label{eq142}
\bar{\rho}^{(j)\prime\prime}_{A\lambda}(\rho_\epsilon^i)=\bar{\rho}^{(j)}_A(\rho^{i\prime\prime}_{\epsilon\lambda}),\\
\bar{\rho}^{(j)\prime\prime}_{B\lambda}(\rho_\epsilon^i)=\bar{\rho}^{(j)}_B(\rho^{i\prime\prime}_{\epsilon\lambda}).
\eeqa
Again since $|\bm{\xi}^i|=|\bm{\xi}^{i\prime\prime}_\lambda|$,  we have
\beq
q^{(j)}(\rho_\epsilon^i)=q^{(i)}(\rho^{i\prime\prime}_{\epsilon\lambda}).
\eeq
Thus, from  theorem one and definition of relative density matrix for mixed states, it follows that
\beq
\label{RRR}
R^{\prime\prime}_\lambda(\rho_\epsilon^{\rm M})=
R(\rho_{\epsilon\lambda}^{\rm M\prime\prime}),
\eeq
where $R^{\prime\prime}_\lambda(\rho_\epsilon^{\rm M})=(A_\lambda\otimes B_\lambda)R(\rho_\epsilon^{\rm M})(A_\lambda^\dagger\otimes B_\lambda^\dagger)/q_\lambda$
is a transformation of the relative density matrix of MRE for $\rho^{\rm M}_\epsilon$, and $R(\rho_{\epsilon\lambda}^{\rm M\prime\prime})$ is a relative density matrix of MRE for the mixed state $\rho_{\epsilon\lambda}^{M\prime\prime}$. It must be emphasized that if there is no any $|\bm{\xi}^i|=0$, the shift $\epsilon$ for the coefficients does not appear, but eqs.(\ref{eq142}-\ref{RRR}) are valid either.  Moreover,  if any component states are maximally entangled, we have to do the replacements such as eqs. (\ref{bellphie}) and (\ref{bellpsie}).  In the last, we take the limitation $\epsilon\rightarrow 0$ to calculate the relative entropy.  Therefore, we obtain that
\beqa
S(\rho_\lambda^{\rm M\prime\prime}\|R^{\prime\prime}_\lambda(\rho^{\rm M}))
&=&\lim_{\epsilon\rightarrow 0}S(\rho_{\epsilon\lambda}^{\rm M\prime\prime}\| R^{\prime\prime}_\lambda(\rho^{\rm M}_\epsilon))\nonumber\\ 
&=&\lim_{\epsilon\rightarrow 0}S(\rho_{\epsilon\lambda}^{\rm M\prime\prime}\| R(\rho_{\epsilon\lambda}^{\rm M\prime\prime}))\nonumber\\ 
&=& S(\rho_\lambda^{\rm M\prime\prime}\|R(\rho_\lambda^{\rm M\prime\prime})),
\label{eq145}
\eeqa
where we have used the fact that the relative entropy is continuous. From monolonicity of relative entropy, it follows that 
\beq
\label{eq146}
S(\rho_\lambda^{\rm M\prime\prime}\|R^{\prime\prime}_\lambda(\rho^{\rm M}))\leq S(\rho^{\rm M}\|R(\rho^{\rm M}))=E_{MRE}(\rho^{\rm M}). 
\eeq
The last equality is because that we have assumed that $R^{\rm M}$ is constructed by the minimum pure state decomposition of $\rho^{\rm M}=\sum_i p_i\rho^i$. 
Again substituting eq.(\ref{eq145}) and the definitions of MRE $\rho_\lambda^{\rm M\prime\prime}$
\beq
E_{MRE}(\rho_\lambda^{\rm M\prime\prime})=\min_{\{p_i,\rho^i\}\in{\cal{D}}} S(\rho_\lambda^{\rm M\prime\prime}\|R(\rho_\lambda^{\rm M\prime\prime}))
\eeq
into (\ref{eq146}), we obtain 
\beq
\label{eq149}
E_{MRE}(\rho_\lambda^{\rm M\prime\prime})
\leq E_{MRE}(\rho^{\rm M}).
\eeq
In terms of joint convexity of relative entropy, we have
\beq
E_{MRE}(\rho^{\rm M}{}^{\prime\prime})\leq\sum_\lambda q_\lambda E_{MRE}(\rho_\lambda^{\rm M}{}^{\prime\prime})
\eeq
Again from eq.(\ref{eq149}) it follows that
\beq
\label{eq150}
E_{MRE}(\rho^{\rm M\prime\prime})
\leq \sum_\lambda q_\lambda E_{MRE}(\rho^{\rm M})
= E_{MRE}(\rho^{\rm M}).
\eeq

It must be emphasized that the precondition that $(A^\dagger_\lambda A_\lambda)\otimes (B^\dagger_\lambda B_\lambda)$ is proportional  to an identity matrix is suggested in order to keep the conservation of probability for the transformed states (which has been normalized) 
\beq
\ket{\psi^{i\prime\prime}_\lambda}_{\rm N}=\frac{1}{\sqrt{q_\lambda q_{i\lambda}}}A_\lambda\otimes B_\lambda\ket{\psi^i}, 
\eeq
and guarantee  the component states $\rho_\lambda^{i\prime\prime}$ with clear significance in the decomposition eq.(\ref{dli}). It is still an open question how to prove  eq. (\ref{eq150}) if there is any $(A^\dagger_\lambda A_\lambda)\otimes (B^\dagger_\lambda B_\lambda)$ that is not proportional  to an identity matrix . 

As to the properties of MRE, in two qubit systems, such as its range is $[0,1]$, its maximum value $1$ corresponds to the maximally entangled states and its minimum value $0$ corresponds to the mixture of the disentangled states, can be directly and easily obtained from the definition of MRE. 
For two qubits, the relative density matrix of MRE is a function of the polarized vectors $\bm{\xi}_A^i,\bm{\xi}_B^i$,  and $\bm{\xi}_A^i,\bm{\xi}_B^i$ are functions of the decomposition density matrices $\rho_i$. Thus, MRE is just a compound function of the decomposition density matrices $\rho_i$. However, in general, a density matrix is not a one to one function of decompositions and a given decomposition is not always able to describe the really physical entanglement.  It is necessary, from our view,  to introduce a new principle so as to determine how to express the measure of entanglement from the a suitable pure state decomposition of density matrix. That is, it seems to us, an  intrinsic physical reason that the requirement of the minimum pure state decomposition is introduced.  Of course, it is not a nice property that a measure of entanglement depends on the possible decompositions because it is not very easy to find all the elements of ${\cal{D}}$. But since the undetermined property of decompositions of the density matrix, it exists in all the known measures of entanglement either. MRE has significantly improvement in this aspect for some kinds of states which has been seen in above section.   We think that it is worth trying to study a thing for any new measure of entanglement.  

\bigskip

In conclusion, MRE can be useful based on six evidences. The first is that MRE is a possible upper bound of DE and a lower bound of EF such as RE, the second is MRE improves the compatibility of RE, the third is that MRE significantly decrease the dependence and sensitivity on the pure state decompositions at least for some interesting states, the fourth is MRE restores the logarithmic dependence from information theory on probability of component states, the five is that MRE reveals the particular difference among some departure states from Bell's states  and the last is that MRE can be extended to multi-party systems naturally\cite{My1}. 

This research is on progressing.

I would like to thank Da Peng Wang for his help to calculate Wootters' EF.

\end{multicols}

\begin{references}
\bibitem{Bennett1}
C.H.Bennett, G.Brassard, C.Cr\'epeau, R.Jozsa, A.Peres, and W.K.Wootters, {\it Phys. Rev. Lett.}{\bf 70},  1895(1993)
\bibitem{Ekert}A. Ekert and R.Jozsa,  {\it Rev. Mod. Phys.} {\bf 68},  733(1996)
\bibitem{QC} D. P. DiVincenzo,  {\it Science} {\bf 270},  255(1995)
\bibitem{QCY}C. A. Fuches,  N.Gisin, R.B.Griffiths, C-S.Niu, and A.Peres,  {\it Phys. Rev. A} {\bf 56},  1163(1997) 
\bibitem{Bennett} C.H.Bennett, H.J.Bernstein, S.Popesu,  and B.Schumacher,  {\it Phys. Rev. A} {\bf 53},  2046(1996);  S.Popescu, D.Rohrlich, {\it Phys. Rev. A} {\bf 56}, R3319(1997)
\bibitem{Vedral1}V.Vedral, M.B.Plenio, K.Jacobs, and P.L.Knight,  {\it Phys. Rev. A} {\bf 56},  4452(1997); V.Vedral, M.B.Plenio, M.A.Rippin, and P.L.Knight,   {\it Phys. Rev. Lett.} {\bf 78},  2275(1997); V.Vedral and M.B.Plenio,  {\it Phys. Rev. A} {\bf 57},  1619(1998)
\bibitem{Wootters}W.K.Wootters, {\it Phys. Rev. Lett.} {\bf 80}, 2245(1998); S.Hill and W.K.Wootters, {\it Phys. Rev. Lett.} {\bf 78}, 5022(1997)
\bibitem{Rains}Eric M. Rains, quant-ph/9809082
\bibitem{Vedral2} V.Vedral, {\it Phys. Lett. A} {\bf 262}, 121(1999) and quant-ph/9903049; M. Muran, M.B.Plenio and V.Vedral, quant-ph/9909031; L.Henderson and V.Vedral, quant-ph/9909011
\bibitem{My1} An Min Wang, quant-ph/0012029
\bibitem{My2} An Min Wang, {\it Chinese Phys. Lett.} {\bf 17}, 243(2000)
\bibitem{Werner} R.F.Werner, {\it Phys. Rev. A} {\bf 40}, 4277(1989)
\bibitem{Peres} A.Peres, {\it Phys. Rev. Lett.} {\bf 77}, 1413(1996)
\bibitem{My4} An Min Wang, quant-ph/0002073
\bibitem{My5} An Min Wang, quant-ph/0011040
\end{references}
\end{document}